\documentclass{emulateapj}

\newcommand{\simgt}{\gtrsim}
\newcommand{\simlt}{\lesssim}

\newcommand{\kms}{\ensuremath{\mathrm{km~s}^{-1}}}

\newcommand{\gcc}{\ensuremath{\mathrm{g~cm^{-3}}}}
\newcommand{\ergg}{\ensuremath{\mathrm{erg~g^{-1}}}}
\newcommand{\ergs}{\ensuremath{\mathrm{erg~s^{-1}}}}

\newcommand{\msun}{\ensuremath{M_\odot}}

\newcommand{\nuc}[2]{\ensuremath{\mathrm{^{#1}#2}}}

\newcommand{\ee}[1]{\ensuremath{\times 10^{#1}}}
\newcommand{\FLASH}{{\sc flash}}
\newcommand{\torch}{{\tt torch47}}
\newcommand{\sPPM}{s{\sc ppm}}

\slugcomment{Submitted to the \apj}

\received{} 
\accepted{} 

\shortauthors{Plewa}
\shorttitle{Dynamics of Detonating Failed Deflagrations}

\begin{document}

\title{Detonating Failed Deflagration Model of Thermonuclear Supernovae I. Explosion Dynamics}

\author{Tomasz Plewa\altaffilmark{1,2}}
\altaffiltext{1}{Center for Astrophysical Thermonuclear Flashes,
   University of Chicago,
   5640 South Ellis Avenue,
   Chicago, IL 60637}
\altaffiltext{2}{Department of Astronomy \& Astrophysics,
   University of Chicago,
   5640 South Ellis Avenue,
   Chicago, IL 60637}
\email{tomek@uchicago.edu}
\begin{abstract} 

We present a detonating failed deflagration model of Type Ia
supernovae. In this model, the thermonuclear explosion of a massive
white dwarf follows an off-center deflagration. We conduct a survey of
asymmetric ignition configurations initiated at various distances from
the stellar center. In all cases studied, we find that only a small
amount of stellar fuel is consumed during deflagration phase, no
explosion is obtained, and the released energy is mostly wasted on
expanding the progenitor. Products of the failed deflagration quickly
reach the stellar surface, polluting and strongly disturbing it. These
disturbances eventually evolve into small and isolated shock-dominated
regions which are rich in fuel. We consider these regions as seeds
capable of forming self-sustained detonations that, ultimately, result
in the thermonuclear supernova explosion.

Preliminary nucleosynthesis results indicate the model supernova
ejecta are typically composed of about $0.1-0.25~\msun$ of silicon
group elements, $0.9-1.2~\msun$ of iron group elements, and are
essentially carbon-free. The ejecta have a composite morphology, are
chemically stratified, and display a modest amount of intrinsic
asymmetry. The innermost layers are slightly egg-shaped with the axis
ratio $\approx 1.2-1.3$ and dominated by the products of silicon
burning. This central region is surrounded by a shell of silicon-group
elements. The outermost layers of ejecta are highly inhomogeneous and
contain products of incomplete oxygen burning with only small
admixture of unburned stellar material. The explosion energies are
$\approx 1.3-1.5\times 10^{51}$~erg.

\end{abstract}
\keywords{hydrodynamics --- nuclear reactions, nucleosynthesis, abundances --- supernovae: general}
\section{Introduction}\label{s:introduction}
Almost half a century ago, \cite{hoyle+60} proposed that some
supernovae may originate from the degenerate remnants of stellar
evolution. These objects are known as Type~Ia supernovae (SN~Ia) and
are commonly believed to be the end points of the evolution of
intermediate mass stars in close binary systems \citep{whelan+73}.

The two most attractive theories of formation of Type~Ia supernovae
are the single-degenerate
\citep[SD;][]{whelan+73,nomoto82,starrfield+04,yoon+04} and double-degenerate
\citep[DD;][]{iben+84,webbink84} scenarios. The observational evidence
necessary to discriminate which formation channel is preferred in
nature remains indirect and fragmentary \citep[and references
therein]{branch+95,livio+03,mannucci+06}, in striking contrast to that
of Type~II supernovae. Evidence supporting the SD scenario has been
collected only recently and in some cases requires more careful
analysis. Some examples include observations of accreting massive
white dwarfs \citep{lanz+05,suleimanov+03}, evidence of hydrogen-rich
material in the vicinity of the supernova, likely associated with a
companion star {hamuy+03}, and the presence of the fast moving
nondegenerate star inside a post-Type~Ia supernova remnant
\citep{ruiz-lapuente+04}, a possible companions of the supernova
progenitor. In the case of the DD formation channel, several close
binary white dwarf systems with total mass around Chandrasekhar mass
have been identified during the last few years
\citep{napiwotzki+03,napiwotzki+05}, essentially confirming double
degenerates as prospective progenitors of thermonuclear supernovae.

Here we limit our considerations to a single-degenerate scenario in
which the ignition of the degenerate matter takes place in the core of
a Chandrasekhar-mass white dwarf. Only a small difference is expected
between the outcomes of the core ignition of a massive white dwarf
formed through either the SD or DD channel. In the latter case, the
explosion is perhaps born in the core of a massive rotating white
dwarf
\citep{piersanti+03}, a remnant of the final merger phase that does
not result in an instantaneous explosion \citep{guerrero+04}.

In defining the initial conditions for multidimensional hydrodynamic
models, we were guided in this study by results of the recent analytic
\citep{garcia-senz+95,wunsch+04,woosley+04} and preliminary numerical
\citep{hoeflich+02,kuhlen+06} studies of conditions prevailing in the
white dwarf's core just prior to the thermonuclear runaway.  Our
limited knowledge of that evolutionary phase grants us certain freedom
in defining starting models. Both the number of the ignition points
and the timing of the ignition are free parameters of the current
models.

Following the failure of carbon \citep{arnett69,arnett+71,arnett74}
and helium detonation models \citep{livne+95}, explosion modeling has
focused on deflagration models and their derivatives. This preference
has been firmly established after the apparent success of parametrized
one-dimensional pure deflagration models
\citep{nomoto+76,nomoto+84}. They, along with the later variant known as a
delayed-detonation model \citep{khokhlov91,woosley90,arnett+94_2dddt}
brought the parametrized models into qualitative agreement with
observations \citep{hoeflich+96}.

Multi-dimensional studies initially included both 2-dimensional models
of deflagrations
\citep{mueller+82_2ddef,mueller+86,livne93_2ddef,livne+93} and
detonations \citep{livne+95}. During the last decade, sophisticated
3-dimensional models have dominated the scene. Central single-point
\citep{khokhlov00,reinecke+02_refined,gamezo+03_3ddef} as well as
multi-point ignition deflagrations
\citep{reinecke+02_3dmpoints,roepke+06_3dmpoints,schmidt+06_3dmpoints,garcia-senz+05_3dmpoints}
seem to produce subluminous events with highly mixed ejecta. Although
models with a parametrized deflagration-to-detonation transition
\citep{gamezo+04,gamezo+05,golombek+05} appear to address both
deficiencies, the mechanism behind the transition to detonation
demands explanation.

There exists evidence \citep[but see also
\cite{blinnikov+06}]{hoeflich+02_1999by,hoeflich+04_2003du,kozma+05}
that the compositional structure of the ejecta obtained in
multi-dimensional centrally-ignited deflagrations may not be
compatible with observations. At the same time, spectroscopic and
polarization observations of several SNe~Ia suggest the existence of
strongly inhomogeneous outer ejecta layers rich in intermediate
elements
\citep{chugai+04_2002ic,wang+04_2004dt,kasen+05,leonard+05}. These two
apparently contradictory requirements can possibly be reconciled
within a class of hybrid models that combine deflagration and
detonation within a single evolutionary sequence. Here we introduce a
detonating failed deflagration (DFD) model, a generalization of our
early exploratory study \citep[hereafter PCL]{plewa+04}, in which both
essential elements are naturally present.  In this model, the
inhomogeneities in the outer ejecta layers result from the large scale
perturbation of the surface stellar layers induced by an off-center
deflagration that fails to unbind the star. In the long term, that
perturbation eventually leads to the formation of isolated
shock-dominated regions that serve as ignition points for a
detonation. The resulting event is luminous with a composite ejecta
structure consisting of smoothly distributed detonation products in
the central regions surrounded by inhomogeneous, turbulent-like outer
layers composed of material partially burned in the deflagration.

Our numerical investigations include certain simplifications with the
assumption of axisymmetry being potentially the most important
one. This, however, will allow us to conduct a small parameter study
exploring the dependence of the explosion parameters on the initial
conditions. In turn, for the first time, we will be able to offer
evidence that the simplifying assumptions regarding the geometry may
not be the major deficiency of our models.
\section{Methods}\label{s:methods}
We study the hydrodynamic evolution of a massive white dwarf using the
\FLASH\ code \citep{fryxell+00}. \FLASH\ has been the subject of rather
extensive verification and validation in both subsonic and supersonic
regimes \citep{calder+02,weirs+05}. We used a customized version of
the code based on the \FLASH{2.4} release with specialized modules
designed to model deflagrations and detonations.  We recorded the
history of individual fluid elements with tracer particles for the
purpose of future detailed nucleosynthesis studies.
\subsection{Reactive hydrodynamics}\label{s:hydro}
We solved the time-dependent reactive Euler equations of
self-gravitating flow in cylindrical geometry assuming axial
symmetry. The non-reactive set of equations were extended by an
advection-diffusion-reaction (ADR) equation describing the evolution
of a deflagration front. The solution to the ADR equation was obtained
with the help of a flame capturing method
\citep{khokhlov95}. The
\FLASH\ implementation has been the subject of verification
\citep{vladimirova+06} with the results of the application to turbulent flames
closely matching the original implementation \citep{zhang+06}.
\subsection{Flame model revision}\label{s:flame}
For the present application, several elements of the original ADR
scheme were modified. In particular, careful analysis of the original
three-stage \FLASH\ burner \citep[PCL]{calder+04} revealed that it
overestimated the amount of energy produced by burning the stellar mix
to nuclear statistical equilibrium (NSE) at densities typical of the
stellar core. As a consequence, the nuclear ashes were both too hot
and too rarefied, with buoyancy effects overestimated by a factor of
$\approx 3$. This led to a much larger acceleration of the burned
material, shorter evolutionary time scales (from ignition to bubble
breakout), a lower consumption of stellar fuel and a correspondingly
lower degree of pollution of the surface layers. The current scheme
captures the energetics of the deflagrating material more closely.  We
also introduced a revised formula for the laminar flame speed that
better approximates the results of \cite{timmes+92}.
\subsubsection{Thermonuclear burning}\label{s:burning}
Following \cite{khokhlov91,khokhlov00}, the evolution of a
deflagrating stellar material can be considered as a sequence of
largely independent processes. Carbon burning (stage I) is followed by
relaxation toward nuclear statistical quasi-equilibrium (NSQE) that
produces silicon-group nuclei (stage II). Eventually, the matter
relaxes toward nuclear statistical equilibrium (NSE) producing
iron-group nuclei (stage III). In the approximate burner, this is
modeled through modifying the composition in all three stages. Stage I
converts carbon into \nuc{24}{Mg} while \nuc{28}{Si} is produced in
stage II by the ``burning'' of \nuc{16}{O} and \nuc{24}{Mg}. We also
extended the original scheme by introducing a ``light nucleus'' that
accounts for the presence of free alpha particles and protons in the
very high density and temperature regime (stage III, see below).

The original \FLASH\ approximate burning scheme assumed the energy
release was a simple sum of the nominal differences between the
binding energy of the initial C/O mix and the burned material, $\Delta
E_b
\approx 7.8\times 10^{17}~\ergg$, independent of density. This is
over 70\% more than the energy release obtained by \citet[Fig.~2,
$\Delta E_b
\approx 4.5\times 10^{17}~\ergg$]{khokhlov83} for the densities of
interest during the initial explosion stages, $\rho\approx 2\times
10^9~\gcc$. Our revised scheme uses results obtained with the \torch\
network \citep{timmes_web} in the self-heating mode. The outcome of
such calculations depends on the initial temperature of the
fuel. Calculations also need to be conducted for long enough to
guarantee a complete burn.

\begin{figure}[t]
\begin{center}
\includegraphics[width=0.45\textwidth,clip=true]{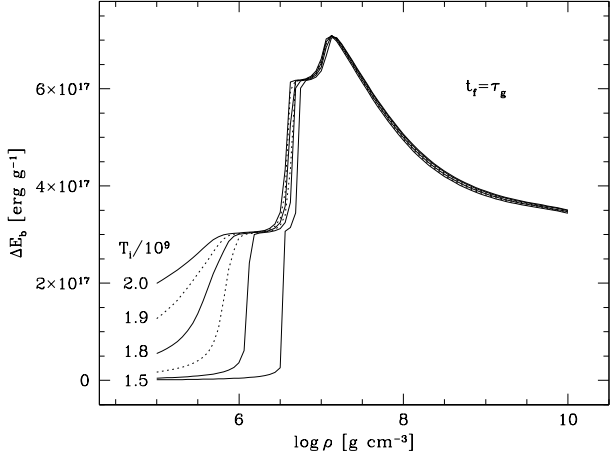}
\caption{Dependence of the energy release on the initial temperature
obtained with \torch\ nuclear network for a 50/50 C/O mixture over a
fixed amount of time, $t_f = \tau_g$. The highest amount of energy is
released for $T_i = 2\times 10^9$~K and the lowest for $T_i =
1.5\times 10^9$~K. The energy release is essentially insensitive to
the initial temperature for densities $\simgt 1\times 10^7~\gcc$.
\label{f:de_rho_ti}}
\end{center}
\end{figure}

In our first set of calculations, we advanced the network for a fixed
amount of time, $t_f = \tau_g$, where $\tau_g = 446 \rho^{-1/2}$~s is
the hydrodynamic (free-fall) time scale \citep{fowler+64,arnett96}.
Calculations were started using several different initial temperatures
of the mixture, $T_i = 1.5-2.0\times 10^9$~K. We found a relatively
weak dependence of the energy release on the initial temperature
(Fig.~\ref{f:de_rho_ti})
%
%
%
%
%
%
and adopted $T_i = 1.7\times 10^9$~K in subsequent calculations.

\begin{figure}[t]
\begin{center}
\includegraphics[width=0.45\textwidth,clip=true]{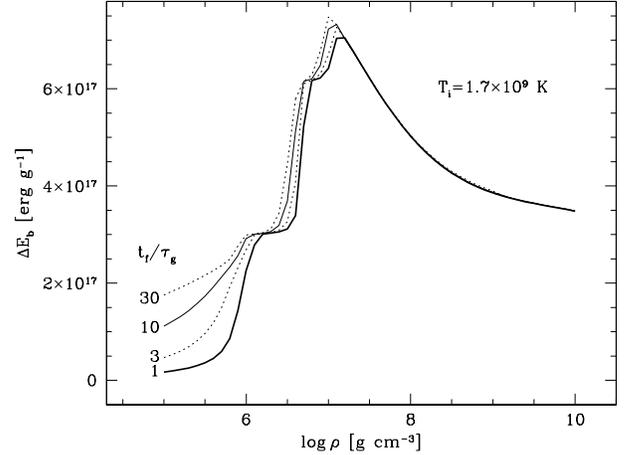}
\caption{Temporal evolution of the energy release in self-heating
\torch\ nuclear network calculations for a 50/50 C/O mix at temperature
 $T_i = 1.7\times 10^9$~K. The bulk of the energy is produced within a
 single hydrodynamic time scale, $\tau_g$. Thick solid lines
 correspond to the energy release adopted in our supernova explosion
 calculations.
\label{f:de_rho_tau}}
\end{center}
\end{figure}

Once the initial temperature was fixed, we turned our attention to
studying the time-dependence of the energy release.  We found that for
$\rho \simgt 1\times 10^6~\gcc$ the bulk of the energy is released
within $\Delta t = \tau_g$ (Fig.~\ref{f:de_rho_tau}).
%
%
%
%
%
%
In the supernova calculations that follow, we used the energy release
obtained for $T_i = 1.7\times 10^9$~K and $t_f = \tau_g$ (thick solid
line in Fig.~\ref{f:de_rho_tau}). The energy release was tabulated as
a function of density with resolution 0.1~dex and linearly
interpolated.  The flame extinction was modeled by smoothly decreasing
the original energy release for densities $< 1\times 10^6~\gcc$. No
energy was released for densities below $1\times 10^5~\gcc$.

\begin{figure}[t]
\begin{center}
\includegraphics[width=0.45\textwidth,clip=true]{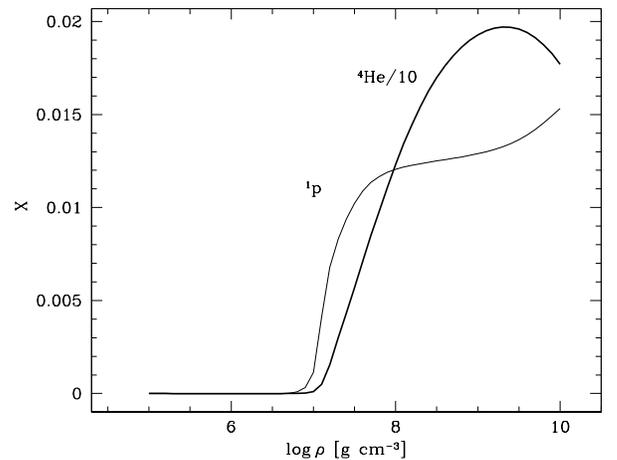}
\caption{Composition of the ``light nuclei'' supplementing the NSE
composition in stage III of the approximate burning scheme.
\label{f:xlight_rho}}
\end{center}
\end{figure}

In contrast to the original \FLASH\ implementation of stage III in
which \nuc{56}{Ni} was the sole product of burning, here we introduced
additional ``light nuclei'' composed of alpha particles and
protons. This allowed us to better approximate the temperature, and in
turn the buoyancy and dynamics of the Rayleigh-Taylor unstable burning
front, especially at early times. As before, we used the \torch\
nuclear network in an isochoric self-heating mode of burning to
determine the composition of the light nuclei
(Fig.~\ref{f:xlight_rho}).
%
%
%
%
%
%
These results appeared insensitive to the particular choice of the
initial temperature or the final time provided the network was evolved
for at least $\tau_g$.
\subsubsection{Coupling to Hydrodynamics}\label{s:flame_coupling}
In addition to tracing the compositional evolution of the nuclear
ashes, the implementation of the Khokhlov's three-stage burner
involves the coupling of the energy source term to hydrodynamics. In
what follows we are primarily concerned with the early evolutionary
stages when the physical time scales associated with nuclear burning
are much shorter than a flame crossing time for computational zones
\citep{khokhlov+93,arnett+94_2dddt,reinecke+02_refined}. When the evolution of the
burning region is relatively slow and the expansion of the star is
insignificant, as it is during the early evolution of centrally
ignited deflagrations, a step-wise form of the energy production rate
leads to large (about two orders of magnitude) rapid fluctuations in
the energy deposition. These localized discrete ``explosions'' occur
in partially-burned material where degeneracy is lifted and are a
source of pressure waves and velocity fluctuations of order $\simeq
10$ km/s. A similar problem in the context of modeling deflagrations
fronts was noted by \cite{khokhlov+93} who used, however, a completely
different procedure to model the flame evolution. The origin of the
problem, i.e.\ discrete representation of a thin unresolved burning
front, is common to both studies.

The impact of the numerical artifact just described can be limited by
appropriately scaling the energy deposition in stage II (no such
procedure is needed in stage I which always operates under strongly
degenerate conditions and no energy is released in stage III). In the
simulations presented here, the energy generation rate for stage II
was kept at 1\% of its nominal value throughout simulations. Given the
advection time scale $\approx 10^{-4}$~s, this procedure affected the
energy release only at very high densities for which the problem of
spurious numerically-induced flow perturbations was originally
identified. Alternatively, one can also consider less intrusive
procedures in which the energy generation rate is limited only for
times $\simlt 0.5$~s. Imposing less strict limitations on the energy
release was found to universally produce velocity perturbations of
magnitude comparable to that of large scale flows.
\subsubsection{Flame speed calibration}\label{s:calibration}
Several factors may cause the numerical flame speed propagation to
differ from the prescribed laminar flame speed in one
dimension. \cite{vladimirova+06} have investigated in some detail the
influence of numerical resolution, evolution with a superimposed
constant background velocity mimicking hydrodynamic advection (i.e.\
to verify Galilean invariance), and the impact of velocity gradients
across the flame front (presumably caused by thermal expansion in real
applications).

In the supernova context, given the degenerate nature of the equation
of state, thermal expansion is controlled by the energy release which
chiefly depends on the fuel density and composition
\citep{timmes+92}. Therefore whenever the flame energetics are
modified, an appropriate calibration of the numerical parameters
controlling the numerical flame speed needs to be done for the
numerical flame speed to match the nominal (laminar) flame speed.

The calibration procedure is relatively simple, but tedious because
the numerical flame speed depends on density, composition, and
numerical resolution. Given our focus is on the evolution of
progenitors composed of a 50/50 carbon/oxygen mix, our flame speed
calibration was limited to that composition. We performed a number of
one-dimensional flame propagation simulations in Cartesian geometry.
Models were obtained on grids with lengths between $1 L$ and $30 L$
with $L=480$~km, and using between $256$ and $1024$ zones. The typical
grid resolution was $\approx 1$~km. The results showed only weak
dependence on the resolution. The flame propagation was, however,
somewhat vulnerable to background flow fluctuations generated by the
flame front motion that freely propagated and partially reflected off
the boundaries. Although we used non-reflecting (zero gradient)
boundary conditions, such reflections are understandable given the
subsonic nature of the problem. In our analysis we used data from
carefully selected long evolutionary sequences computed on the largest
grids and at the highest possible resolution.

Our calibration procedure was applied toward a slightly modified
formula originally proposed by \citet[Eq.~43]{timmes+92}:
\[
s_{l,TW} = a {\rho}_{9}^{b} \times \left(1.001 - e^{-\left(\frac{\rho_9-\delta_q}{\sigma_q)}\right)^2}\right)
\]
with $a=35.46538\times 10^5$, $b=1.110603$, $\delta_q =
2.6132427\times 10^{-2}$, and $\sigma_q = 2.9538546\times
10^{-2}$. This improved formula reproduces \cite{timmes+92}
results for a C/O 50/50 mix to within 15\% for densities between
$1\times 10^7$ and $1\times 10^9~\gcc$. Finally, this has been
adjusted to account for thermal expansion due to the \torch\
energetics:
\[
s_{l,FLASH} = s_{l,TW} \times \max(0.9, \min(1.3, p(\log_{10}\rho) )),
\]
where
\[
p(x) = c_0 + c_1 x + c_2 x^2 + c_3 x^3 + c_4 x^4,
\]
is polynomial density-dependent correction factor based on our
calibration calculations and $c_0=413.6563$, $c_1=-194.3208$,
$c_2=34.06912$, $c_3=-2.633218$, and $c_4=7.5665459\times
10^{-2}$. The final formula reproduces \citet[Table~3]{timmes+92}
results to within 5\% for densities in the range $1\times 10^7$ and
$4\times 10^9~\gcc$.
\subsubsection{Hybrid burning scheme}\label{s:hybrid}
Our initial investigations into late evolutionary stages of fizzle
off-center deflagrations (PCL) provided strong evidence that the
conditions inside the confluence region at the stellar surface are
appropriate for creating a detonation. We observed that both densities
and temperatures were high enough for the burning time scale to become
shorter than hydrodynamic time scale. We also anticipated that the
shocked region would remain confined for an extended period of time
sufficient to develop a self-sustained detonation.  We were unable,
however, to study that process in detail at that time.  The burning
module could not reliably discriminate between shocked fuel (a
legitimate detonation site) and compressed partially burned
matter. Here we introduce a hybrid burning scheme to allow for the
simultaneous presence of deflagrations and distributed nuclear burning
in the simulation. Deflagrations are modeled using the ADR
flame-capturing scheme \citep{khokhlov95,vladimirova+06} with
modifications as outlined above.  The distributed burning is
calculated using the \FLASH\ {\tt aprox13t} 13-isotope alpha network
(G.~Jordan 2005, private communication) This network is an extension
of the original \FLASH\ {\tt approx13} network and includes
temperature coupling for increased stability of calculations in the
NSE regime \cite{mueller86}.

The first step in our our hybrid burning procedure is to identify
shocked zones. This is done using a multidimensional shock detection
module adopted from the \sPPM\ code \citep{anderson+95} with pressure
jump across the shock, $\Delta p/p \geq 0.5$. If the temperature in
such zones is high enough for nuclear burning ($T_{nuc,min} = 8\times
10^8$~K), a progress variable used by the ADR flame module is
reinitialized and the numerical flame speed is set to zero. The flame
module is subsequently activated only if the flame speed is greater
than zero and the material is pure fuel (as indicated by, for example,
a small abundance of nuclei not participating in the simplified
three-stage burner). Otherwise, the nuclear network is called but only
if the following conditions are fulfilled simultaneously: (1) both the
density and temperature are in the range appropriate for the network
calculations, (2) the zone is outside a shock front, (3) the
deflagration module was not used, and (4) the ADR progress variable is
small. The first condition limits network calculations to high-density
($\rho \ge 1\times 10^5~\gcc$) high-temperature ($T \ge 8\times 10^8$
K) regimes. Condition (2) follows from the conclusions of
\cite{fryxell+89} who found that the correct speed of detonation waves can
be obtained only in models with burning disallowed inside unrelaxed
numerical shock profiles. Conditions (3) ensures only one physical
burning process, either distributed burning following shock heating or
deflagration, operates at a time. Finally, condition (4) prevents
burning from occurring in regions preheated by an extended diffusive
tail of the ADR flame capturing scheme and at the same time allows for
burning to continue after the detonation wave sweeps through partially
deflagrated material. By making the progress variable threshold for
use of nuclear network slowly increasing with time, we also prevented
spurious (distributed) burning in the dense central regions perturbed
by ascending deflagrating material.

The above scheme captures the essential behavior of both deflagrations
and wave-induced distributed burning. It allows for the transition to
detonation in shock-heated regions away from deflagrating material and
enables burning in shocked, partially deflagrated material. The above
selection rules were developed and improved based upon experience
accumulated in the course of several numerical experiments and, as
such, offers a practical rather than ideal recipe for modeling
deflagrations and detonations in the same physical setting including a
possible deflagration to detonation transition.

Once the detonation wave is launched, a combination of a PPM
hydrodynamic solver, a multidimensional shock detection algorithm, and
a {\tt aprox13t} nuclear burning module was used to advance the
evolution in time.  The evolution of the detonation wave on a large
scale is expected to be captured correctly given that the thickness of
the wave is much smaller than the stellar radius \citep{falle00}.  We
did not find it necessary to rescale nuclear reaction rates in order
to obtain the correct propagation of the detonation speeds
\citep{arnett+94_2dddt}; excluding the unrelaxed shock profile from
burning appeared sufficient to yield physical solutions.
\subsection{Some comments on numerical model limitations}\label{s:numerics}
One of primary motivations behind development of a hybrid burning
scheme was desire to mitigate a risk of spurious detonation ignitions.
The existence of such spurious ignitions is a well-documented fact in
astrophysical literature \citep{fryxell+89}. Here we only briefly
discuss the most typical causes for spurious ignitions and possible
ways to prevent them from polluting hydrodynamical models.

Perhaps the most common cause for spurious detonation ignitions is a
numerically-induced mixing of hot ashes and cold fuel.  By advecting a
material interface (contact discontinuity) separating ashes from fuel,
\cite{fryxell+89} demonstrated that such mixing may result in
artificial preheating of fuel, its ignition, and ultimately formation
of a combustion wave.  Related to this is a problem of species
conservation by nonlinear Eulerian advection schemes. Under certain
conditions numerical modification of fuel (or partially burned
material) composition may change burning energetics. Either extinction
or enhancement of burning may follow. To our knowledge, neither
numerical species diffusion nor species non-conservation can be
completely eliminated from Eulerian simulations of realistic nonlinear
systems.  Our hybrid burning scheme attempts to limit possible
influence of both effects by constraining burning to regions occupied
by pure fuel.

Transition to detonation can also follow an artificial boosting of an
acoustic perturbation.  Such perturbations lead to local variations in
temperature and under degenerate conditions temperature is a sensitive
function of (relatively small thermal) pressure. Given strong
dependence of nuclear reactions on temperature, it is conceivable that
even small but sustained heating may strengthen acoustic waves and
eventually cause spurious transition to detonations. Typically,
however, small acoustic fluctuations suffer from strong damping by
numerical diffusivity and such spurious transitions to detonations are
likely to occur only if nuclear burning is allowed inside numerical
(unrelaxed) shock profiles. The cure for this problem is to eliminate
burning in regions occupied by shocks \citep{fryxell+89} and, as we
mentioned above, such a filter is employed in our calculations.

Finally, application of nuclear burning source term in our
calculations is limited to regions with sufficiently high densities
and temperatures.  That is, we wish to consider a feedback from
nuclear burning only if the nuclear timescale is short enough to
influence hydrodynamics. This approach not only saves computational
time, but more importantly prevents the nuclear network from being fed
with input data representing low-density regions where evolution has a
highly transient character and is not correctly captured in our
calculations.

To summarize, numerical computations of reactive flows pose extreme
challenges and require very careful treatment. We attempted to address
several known and some newly emerged problems related to coupling
reactive source terms to hydrodynamics in great detail. The impact of
some of these problems could only be limited, but not completely
eliminated. For example, in our models no nuclear burning is allowed
inside numerical shock profiles. However, our particular choice of
parameters defining numerical shock profile may not be adequate in all
situations affecting evolution of shocks and acoustics in unwanted
manner. Poor numerical resolution only adds to the algorithmic
inefficiency further limiting predictive abilities of our models.

Clearly, successfully resolving technical problems of our computations
is of high priority and such aspects should be remembered when
interpreting our results. At the same time, however, one should also
keep in mind that our model involves several simplifying assumptions
(i.e., we consider a non-rotating, non-convective, non-magnetized
chemically homogeneous progenitor), and as such it is a blend of
approximate numerics and unique choice of parameters defining physical
scenario.
\subsection{SN~Ia explosion code verification: central deflagrations}\label{s:central}
Given that both the basic hydrodynamic module as well as the FLASH
implementation of the flame capturing scheme have been extensively
verified and, albeit to a limited extent, also validated in the past
\citep{calder+02,weirs+05,vladimirova+06}, our verification is solely
limited to code-to-code comparison in the context of thermonuclear
supernova explosion modeling. Although code-to-code comparison is
widely popular among computational physicists, the usefulness of this
approach is hotly debated \citep{trucano+03}. For one, even perfect
agreement between the simulation results of two codes does not offer
proof of their correctness. Moreover, the scope of such an exercise is
usually limited by the specific capabilities of the code, the
availability of the results, the completeness of documentation, and
the relevance of the performed test to the actual problem at hand, to
name a few. Here we accept the obvious deficiencies of the
code-to-code comparison approach and use highly relevant and
state-of-the-art calculations as a benchmark. Again, due to the
novelty of our ultimate application, no data are available for
comparison, though they will hopefully become available through
independent calculations. Ultimately, the model will be validated
using accumulated observational evidence. Methodology and methods
required in such assessments are presented in a companion paper
\citep[][see also
\cite{blinnikov+06}]{kasen+06}.
\subsubsection{Simulation setup}\label{s:central_setup}
For our comparison study we selected a family of centrally ignited
deflagration models obtained by the Garching group \citep[and
references therein]{reinecke+02_refined,roepke+05_full}. These studies
are very well-documented and their results compare favorably to those
obtained by other groups \citep{gamezo+03_3ddef,gamezo+04}. When
comparing results, we considered the overall morphology of the
explosion (flame front structure), the energetics and the approximate
ejecta composition of the 2-dimensional explosion calculations
reported by
\cite{reinecke+02_refined} and \cite{roepke+05_distributed}. Our
choice of 2-D Garching models is natural given our calculations to
follow also assume axial symmetry.

In a benchmark study we used \FLASH{2.4} customized for the
thermonuclear supernova explosion problem. We used a PPM module for
real gas inviscid hydrodynamics and the Helmholtz equation of state
required by the degenerate conditions encountered in the white dwarf
interior. All calculations were done with Courant factor $0.6$. This
choice of time step limiter may appear somewhat conservative, but
allows for better coupling between different physical processes.

Gas self-gravity was accounted for by solving the Poisson equation
through multipole expansion. We found that linear momentum is poorly
conserved in explosion calculations when the expansion series is
truncated too early, especially when the explosion displays
significant asymmetries. In test calculations done assuming a
spherical potential, the bulk of the stellar material displayed motion
of a few hundred km/s after only a few seconds of evolution. The
momentum conservation gradually improved as additional higher order
terms were included in the expansion. In what follows we used 10
multipole moments and found excellent momentum conservation for all
initial conditions considered.

We used a 2-dimensional cylindrical grid $(r,z)$ and assumed axial
symmetry. This implied imposing a reflecting boundary condition at
$r_{min}=0$. We applied outflow conditions at the remaining
boundaries. In our verification calculations the computational domain
covered a rectangular region with $r_{max}=z_{max}=-z_{min}=16,384$
km. We used the adaptive capabilities of the \FLASH\ code to create
several levels of refinement up to a maximum resolution of 8~km. We do
not expect the dynamical evolution of low density gas or at large
distances from the stellar center to play an important role in
explosion simulations and therefore adaptive refinement was used only
for radii $< 4,000$~km and if the gas density $> 1\times 10^4~\gcc$.
Self-gravity calculations, on the other hand, require good resolution
of dense regions and the grid refinement was forced to the highest
allowed resolution whenever the gas density $> 3\times 10^6~\gcc$.  In
addition, the innermost $2,500$~km of the star have always been
resolved with the finest zones. In regions where adaptivity was
allowed, AMR patches were created dynamically if the local density
contrast exceeded $0.50$ or the total velocity changed by more than
20\%.  Furthermore, we ensured the flame front was always refined to
the finest level by forcing refinement whenever the fractional change
of the flame progress variable exceeded $0.02$.
\subsubsection{Initial model}\label{s:central_initial}
The supernova progenitor was an isothermal, $T = 5\times 10^7$~K,
white dwarf composed of equal mass fractions of carbon and
oxygen. With a central density of $2\times 10^9~\gcc$, the progenitor
had a radius $\approx 2100$~km, total mass $\approx 1.36~\msun$, and
total energy $\approx -4.92\times 10^{50}$ erg. The progenitor was
surrounded by a low density ($\rho_{amb} = 1\times 10^{-3}~\gcc$) and
low temperature ($T_{amb} = 3\times 10^7$~K) medium composed of pure
helium. The stellar material and low density ambient medium were
initially marked with a passively advected tracer that was set to $1$
and $0$ in those two regions, respectively. Subsequently, gas
gravitational accelerations were multiplied by the tracer value in the
course of the evolution, allowing us to prevent the ambient medium from
collapsing onto the central object and limiting mass diffusion at the
stellar surface.

The original progenitor model was constructed using a numerical
discretization different from the one used in the hydrodynamic
simulations and assuming a simplified equation of state. For that
reason, the perfect hydrostatic equilibrium of the original model was
destroyed as soon as it was interpolated onto the simulation
mesh. Even though the mismatch between the two computational
environments was relatively small, very strong acoustic oscillations
quickly developed making such a model unsuitable for further
investigations. Moreover, the oscillations did not decay with time,
presumably due to both low dissipation of the hydrodynamic scheme and
the strong degeneracy of the medium.

\begin{figure}[t]
\begin{center}
\includegraphics[width=0.45\textwidth,clip=true]{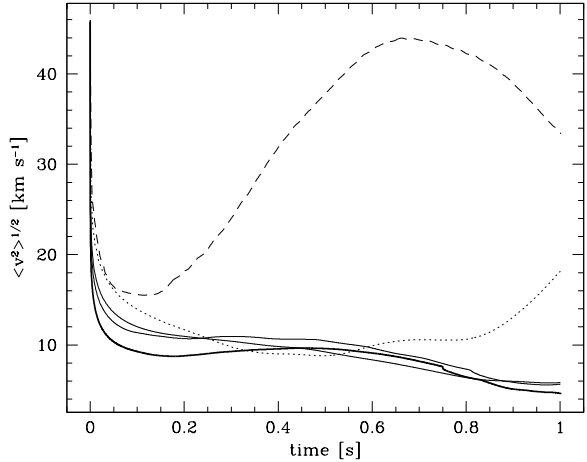}
\caption{Evolution of the root mean square velocity in models without
  burning obtained with resolution 2 (thick solid), 4 (medium solid),
  8 (thin solid), 16 (dotted), and $32$~km (dashed). The initial
  models were perturbed by adding random velocities with amplitude
  $200~\kms$ inside the stellar core. Note that the velocity gently
  decays in models calculated with a resolution of $16$~km or better.
\label{f:static_vrms}}
\end{center}
\end{figure}

\begin{figure}[t]
\begin{center}
\includegraphics[width=0.45\textwidth,clip=true]{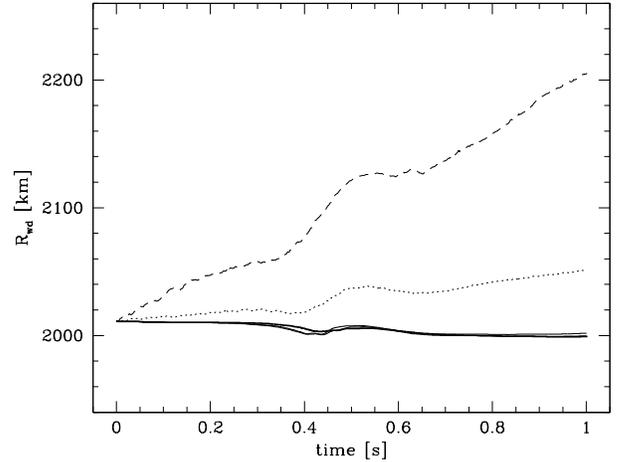}
\caption{Evolution of the stellar radius in models without burning
  obtained with resolution 2 (thick solid), 4 (medium solid), 8 (thin
  solid), 16 (dotted), and $32$~km (dashed). The initial models were
  perturbed by adding random velocities with amplitude $200~\kms$
  inside the stellar core. Models computed with resolution no worse
  than $8$~km are stable, displaying only a very small degree of
  radius change. The model using $32$~km resolution shows very strong
  expansion, making it unsuitable for long-term hydrodynamic evolution
  studies.
\label{f:static_radius}}
\end{center}
\end{figure}

We constructed a stable progenitor model using a modified variant of
the damping method of \cite{arnett94} in a 1-D \FLASH\ simulation in
spherical geometry. Our procedure combined a very slow diffusion of
velocity together with a partial rather than complete (as in the
original method) removal of excess momentum after each time step. We
found that the complete removal of momentum prevented the model from
reaching equilibrium. Damping process usually lasted several thousands
of hydrodynamic steps. We examined the stability of the relaxed model
by computing a sequence of hydrodynamic models without burning.
Random velocity perturbations with amplitude of $200~\kms$
($\mathrm{Ma}\approx 0.02$) were added to the inner core region of
radius $400$~km. We observed the decay of the root mean square
velocity with time, as expected in a stable model, provided the
resolution was $16$~km or better (Fig.~\ref{f:static_vrms}).
%
%
%
%
%
%
We also computed the evolution of an ``effective'' stellar radius
corresponding to the volume occupied by gas with density $> 1\times
10^4~\gcc$ representing the bulk of the stellar matter. The results
are shown in Fig.~\ref{f:static_radius}.
%
%
%
%
%
%
As one can see, the model stellar radius shows significant evolution
in computations with resolution $16$~km or worse; only a very modest
($\approx 0.5\%$) decrease of radius was observed in the better
resolved models, and had no consequence for structure of the stellar
core (i.e.\ possible temperature increase).

\begin{figure}[t]
\begin{center}
\includegraphics[width=0.45\textwidth,clip=true]{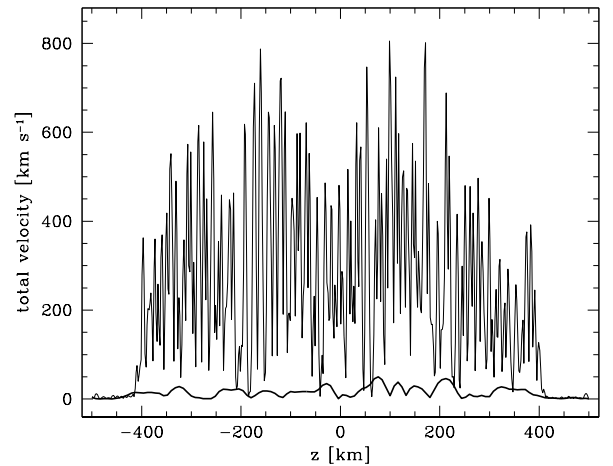}
\caption{Total velocity profiles at $t=0.1$~s near the symmetry axis
  ($r=0$~cm) in models without burning computed with $2$~km (thick
  line) and $8$~km (thin line) resolution. The velocity amplitude
  rapidly and significantly increases in the $2$~km model, while it
  slowly decreases in the $8$~km model.
\label{f:static_vaxis}}
\end{center}
\end{figure}

Unfortunately, close examination of the velocity field also revealed
that while the overall stability of models improves with increasing
resolution, small velocity perturbations not only do not decay but are
amplified near the symmetry axis. This effect was especially strong in
$2$~km resolution model where the velocity near the axis rapidly
increased from the initially imposed $200~\kms$ to $800~\kms$ during
the first $0.1$~s of the evolution (shown with the thin solid line in
Fig.~\ref{f:static_vaxis}).
%
%
%
%
%
%
No significant increase in the magnitude of the spurious velocities
was observed at later times, but the affected region expanded from the
initial 1 to 3 zones by $t=1$~s.  We observed very similar behavior in
the $4$~km resolution model, although the velocities near the symmetry
axis were ultimately somewhat smaller ($\approx 600~\kms$). In
contrast, the velocities smoothly decay from their initial values in
the whole perturbed region in the $8$~km resolution model (shown with
the thick solid line in Fig.~\ref{f:static_vaxis}). A similar slow
decrease of the velocities near the symmetry axis was also observed in
the $16$~km model.

After a stabilized progenitor model was interpolated onto the
simulation mesh, a deflagration front was initialized around a small
region at the stellar center. The flame front was located at
\[
R_{f} = R_{m} \times \left[1 + a_{m} \cos\left(n_{m} \tan^{-1}\left(\frac{z}{r}\right)\right)\right]\,,
\]
where $R_{m}$ is the unperturbed radius of the burned region (flame
radius), $a_{m}$ is the flame radius perturbation amplitude, and
$n_{n}$ controls the flame radius perturbation wavelength. Random
velocity perturbation were added to the inner core region following
the procedure described above. By introducing a small stochastic
component (``cosmic variance'') into the problem, we were able to
examine the robustness of our results. In particular, it is essential
to verify that the observables (i.e.\ the explosion energy) do not
show strong dependence on small perturbations in the initial
conditions, in accord with the observed intrinsic homogeneity of SN~Ia.
\subsubsection{Centrally ignited benchmark models}\label{s:central_database}
\begin{figure*}[ht]
\begin{center}
\includegraphics[height=5.5cm,clip=true]{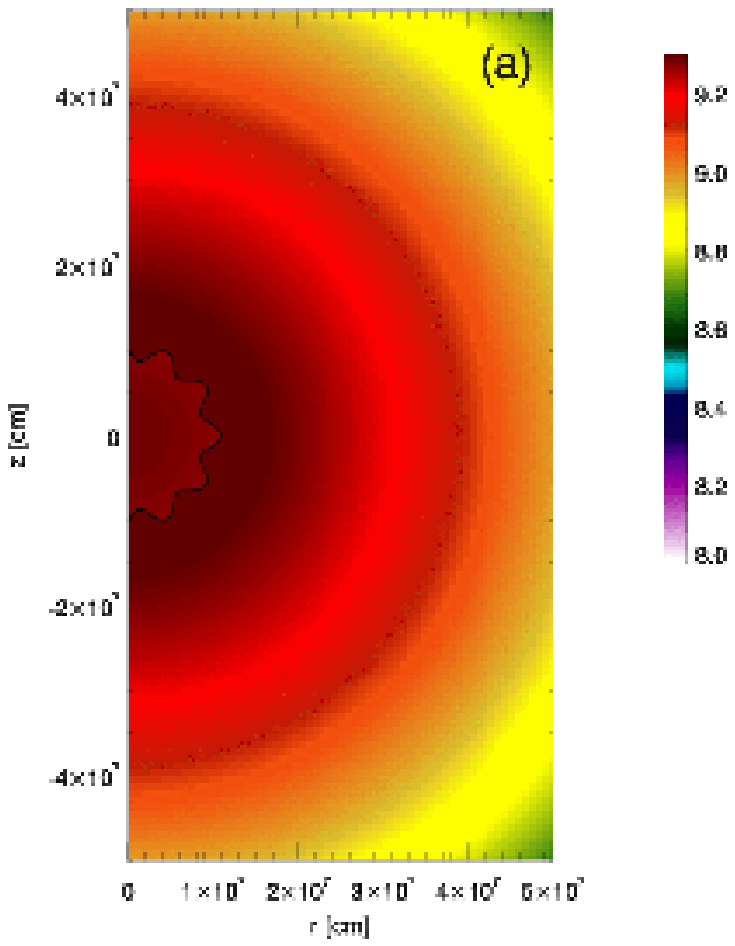}%
\includegraphics[height=5.5cm,clip=true]{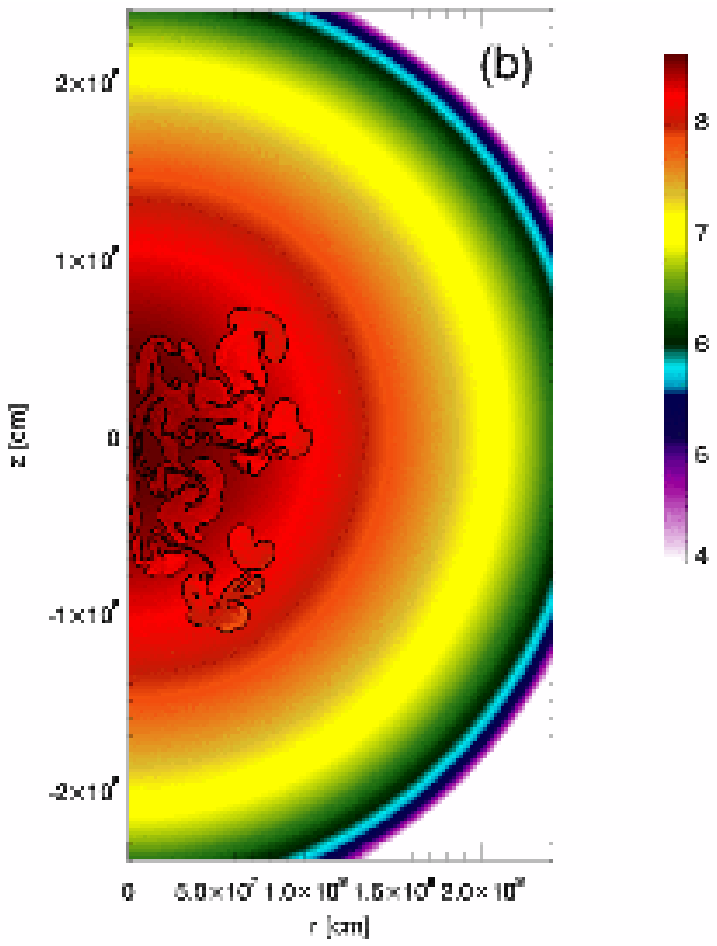}%
\includegraphics[height=5.5cm,clip=true]{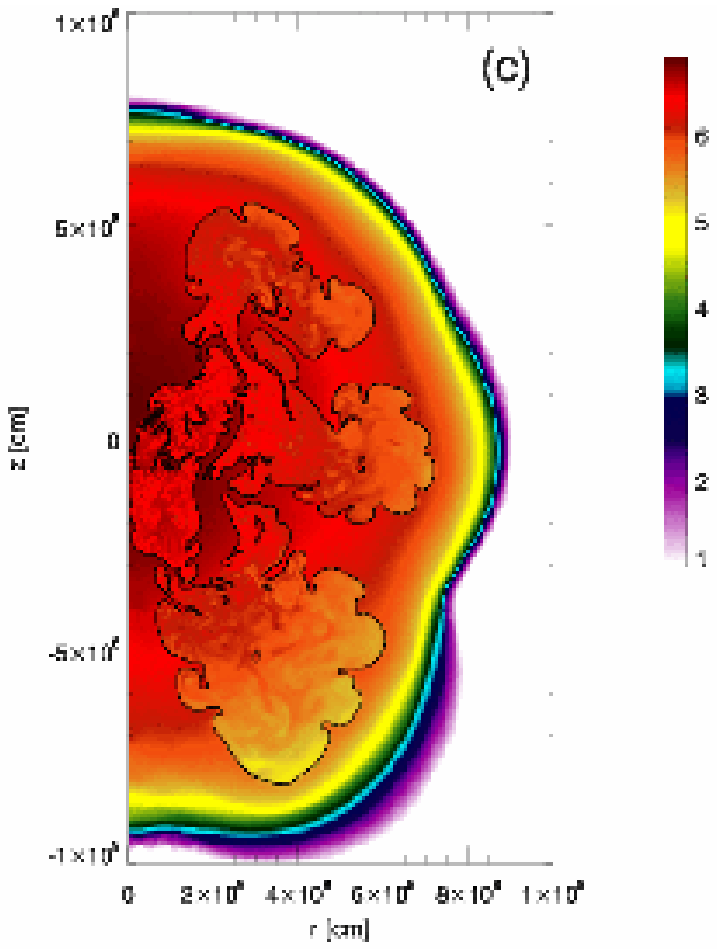}%
\includegraphics[height=5.5cm,clip=true]{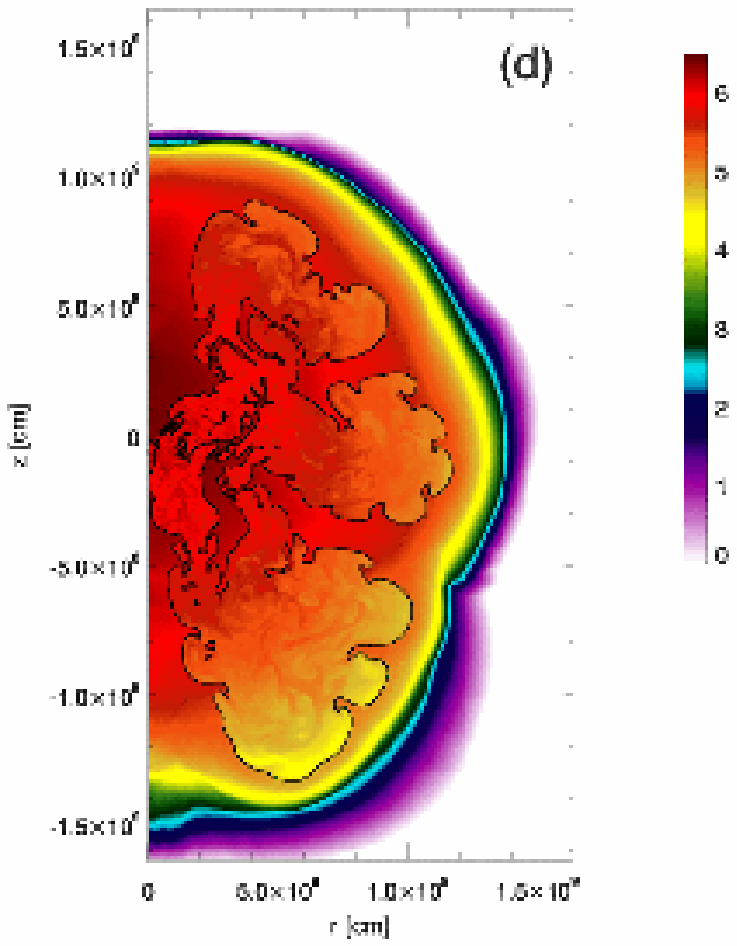}
\end{center}
\caption{Hydrodynamic evolution of the centrally ignited benchmark
  supernova model n11d2r10t10a. Panels (a)-(d) show the density in
  logarithmic scale together with the outline of the flame front
  (contour line corresponding to progress variable value of
  $0.5$). Notice the scale change between panels. (a) The initial
  conditions at $t=0$~s. (b) $t=1.0$~s. The flame front is highly
  convoluted; the star remains spherical but expanded by $\approx
  20$\%; the bulk expansion velocity is $\approx 2,500~\kms$. (c)
  $t=2.0$~s. The flame front is rich in structure with some pockets of
  unburned material; a large amount of unburned material can be found
  near the center; individual flame bubbles create smooth large scale
  impressions on the stellar surface; the expansion velocity near the
  outer stellar edge are just slightly over $10,000~\kms$. (d)
  $t=2.5$~s. The flame is extinct and the structure of the flame front
  becomes frozen. Evolution toward homologous expansion begins.}
\label{f:central_time}
\end{figure*}

We performed a comprehensive survey of two-dimensional centrally
ignited deflagration models. The database contains 33 models. All
models were evolved until $t=2.5$~s when burning essentially
ceased. For each model we varied the flame perturbation wavelength and
random velocity perturbation pattern (through the seed
perturbation). We explored the sensitivity of the results to mesh
resolution by resolving the central stellar region of radius $R_{c}$ t
${\Delta}x_{c}$ for times $< t_{c}$. Resolution coarsening in this
central region was done in equal intervals of time. For times $>
t_{c}$, the resolution was equal to a default value of $8$~km. The
initial flame radius and perturbation amplitude were in all cases
fixed at $R_m=100$~km and $0.1$, respectively. Table~\ref{t:central}
%
%
%
\begin{table}
\caption{Centrally Ignited Benchmark Deflagration Models}\label{t:central} 

\begin{tabular}{lrrrrr}
Model         &  $n_m$  &    ${\Delta}x_c$\ [km]  &  $R_{c}$\ [km] & $t_{c}$\ [s] & $E_t$\tablenotemark{a}\ [$10^{51}$ erg]\\
\tableline
n7d1r10t15b   &    7    &      1                 &    1000         &   1.5        &        0.86            \\
n7d1r10t15c   &    7    &      1                 &    1000         &   1.5        &        0.97            \\
\tableline
n9d1r10t15b   &    9    &      1                 &    1000         &   1.5        &        0.75            \\
n9d1r10t15c   &    9    &      1                 &    1000         &   1.5        &        0.66            \\
\tableline
n11d1r10t15b  &   11    &      1                 &    1000         &   1.5         &        0.75           \\
n11d1r10t15c  &   11    &      1                 &    1000         &   1.5         &        0.66           \\
\tableline
n11d2r05t10a  &   11    &      2                 &     500         &   1.0         &        0.27           \\
n11d2r05t10b  &   11    &      2                 &     500         &   1.0         &        0.38           \\
n11d2r05t10c  &   11    &      2                 &     500         &   1.0         &        0.44           \\
n11d2r05t10d  &   11    &      2                 &     500         &   1.0         &        0.38           \\
n11d2r05t10e  &   11    &      2                 &     500         &   1.0         &        0.38           \\
\tableline
n11d2r10t10a  &   11    &      2                 &    1000         &   1.0         &        0.35           \\
n11d2r10t10b  &   11    &      2                 &    1000         &   1.0         &        0.36           \\
n11d2r10t10c  &   11    &      2                 &    1000         &   1.0         &        0.42           \\
n11d2r10t10d  &   11    &      2                 &    1000         &   1.0         &        0.38           \\
n11d2r10t10e  &   11    &      2                 &    1000         &   1.0         &        0.39           \\
n11d2r10t10f  &   11    &      2                 &    1000         &   1.0         &        0.36           \\
\tableline
n11d2r10t15a  &   11    &      2                 &    1000         &   1.5         &        0.56           \\
n11d2r10t15b  &   11    &      2                 &    1000         &   1.5         &        0.64           \\
n11d2r10t15c  &   11    &      2                 &    1000         &   1.5         &        0.60           \\
n11d2r10t15d  &   11    &      2                 &    1000         &   1.5         &        0.52           \\
n11d2r10t15e  &   11    &      2                 &    1000         &   1.5         &        0.58           \\
\tableline
n11d2r20t15b  &   11    &      2                 &    2000         &   1.5         &        0.66           \\
n11d2r20t20b  &   11    &      2                 &    2000         &   2.0         &        0.54           \\
\tableline
n11d2r40t20b  &   11    &      2                 &    4000         &   2.0         &        0.54           \\
n11d2r40t20b3 &   11    &      2                 &    4000         &   2.0         &        0.55           \\
n11d2r40t20b6 &   11    &      2                 &    4000         &   2.0         &        0.52           \\
\tableline
n11d2r40t15b  &   11    &      2                 &    4000         &   1.5         &        0.79           \\
n11d2r60t15b  &   11    &      2                 &    6000         &   1.5         &        0.79           \\
\tableline
n13d1r10t15b  &   13    &      1                 &    1000         &   1.5         &        0.57           \\
n13d1r10t15c  &   13    &      1                 &    1000         &   1.5         &        0.69           \\
\tableline
n15d1r10t15b  &   15    &      1                 &    1000         &   1.5         &        0.57           \\
n15d1r10t15c  &   15    &      1                 &    1000         &   1.5         &        0.52           \\

\end{tabular}
\tablenotetext{a}{Total energy at $t=2.5$~s.}
\end{table}
%
%
%
presents a complete database of our benchmark models.\footnote{The
database is available on-line at {\tt
flash.uchicago.edu/\~tomek/Results/DFD/central/}.}  The model
identification tag name is constructed as a string {\tt n--d-r--t--ph}
where {\tt n--} denotes the perturbation pattern, {\tt d-} describes
the maximal grid resolution in the core region, {\tt r--} denotes a
radius of the core region inside which enhanced resolution was used,
{\tt t--} denotes the time up to which enhanced flame resolution was
allowed, {\tt p} distinguishes between different perturbation
patterns, and finally {\tt h} denotes the desired thickness of the
flame front (in grid cells) in the ADR flame capturing module
\citep[see Appendix in][]{khokhlov95}. The default value of the numerical
flame thickness was 4; we obtained two models, n11d2r40t30b3 and
n11d2r40t30b6, in which the nominal flame thickness was varied by
$-25$ and $+50$\%, respectively. The last column in
Table~\ref{t:central} gives the total energy of the model (sum of
kinetic, internal, and potential energies) at the final time,
$t=2.5$~s.
\paragraph{Evolution in reference model}
We use model n11d2r10t10a to illustrate major characteristics of a
centrally ignited deflagration in our benchmark configuration
(Fig.~\ref{f:central_time}).
%
%
%
%
%
%
The density in this model at the initial time is shown in
Fig.~\ref{f:central_time}(a). The flame front has the shape of the
initial regular $n=11$ perturbation. At a resolution of $2$~km,
comparable to that of the highest resolution Garching group models at
early times
\citep[Fig.~3]{roepke+06_3dmpoints}, the flame region is resolved into
about 50 zones in radius. This initial configuration undergoes a
dramatic evolution during the next second
(Fig.~\ref{f:central_time}(b)). After that time only 3 prominent
bubbles are clearly identifiable and some parts of the flame begins
forming disconnected regions (e.g.\ the region located near $(r,z)
\approx (650, -550)$~km).  The flame leaves behind a significant
amount of unburned material and a few isolated pockets of fuel can
also be identified inside rising bubbles.  The outermost parts of a
highly convoluted flame front have reached $\approx 1,000$~km in
radius.  The star has expanded by $\approx 20$\% and the typical
expansion velocities are $\approx 2,500~\kms$. After another second
($t=2$~s; Fig.~\ref{f:central_time}(c)), the expansion of the outer
stellar layers becomes slightly nonuniform due to the uneven
acceleration caused by individual flame bubbles. The expansion
velocity exceeds $10,000~\kms$ near the stellar surface. At this time
one can still identify 3 large bubbles, but these are now more
developed and occupy a much larger volume fraction. Their morphology
does not change much at still later times ($t=2.5$~s, the final time;
Fig.~\ref{f:central_time}(d)) as nuclear burning essentially quenches
and the ejecta expansion becomes progressively radial.
\paragraph{Sensitivity to small perturbations}
Given the highly nonlinear character of the Rayleigh-Taylor unstable
deflagrating bubbles it is natural to ask how robust are the
predictions offered by individual models. We studied this question by
creating a sequence of models for a given set of primary model
parameters (initial flame configuration and numerical resolution). The
initial conditions for each member of a given sequence differed only
by the pattern of small amplitude stochastic velocity added to the
initially static progenitor model. Our metric for comparison is to
examine the final ejecta morphology and integral model characteristics
(temporal evolution of total energy, burning rate, flame surface
area).

\begin{figure*}[ht]
\begin{center}
\includegraphics[height=5.5cm,clip=true]{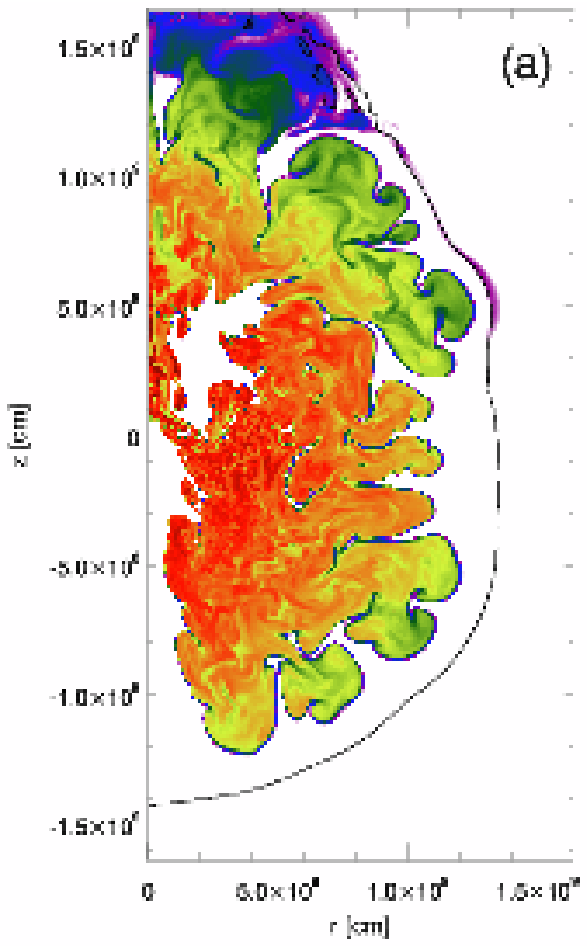}%
\includegraphics[height=5.5cm,clip=true]{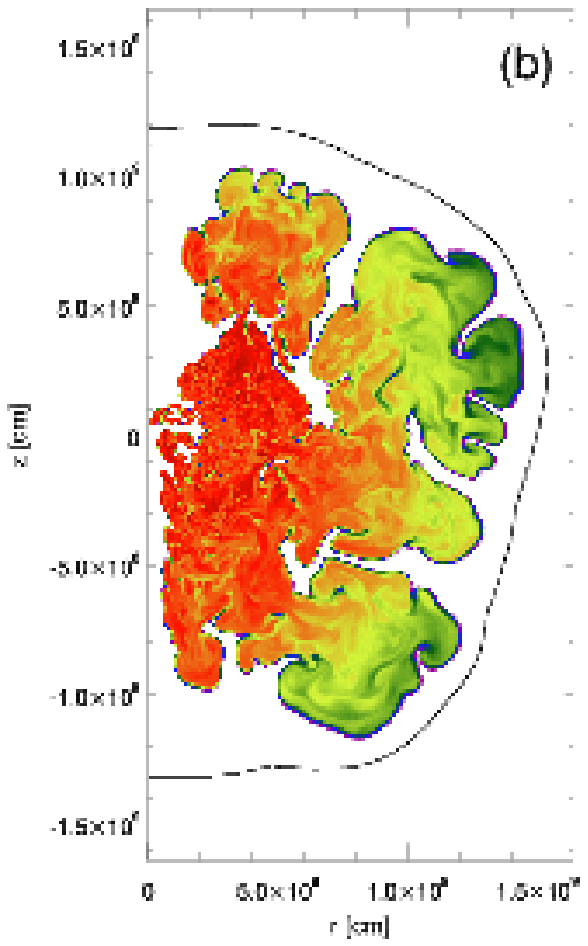}%
\includegraphics[height=5.5cm,clip=true]{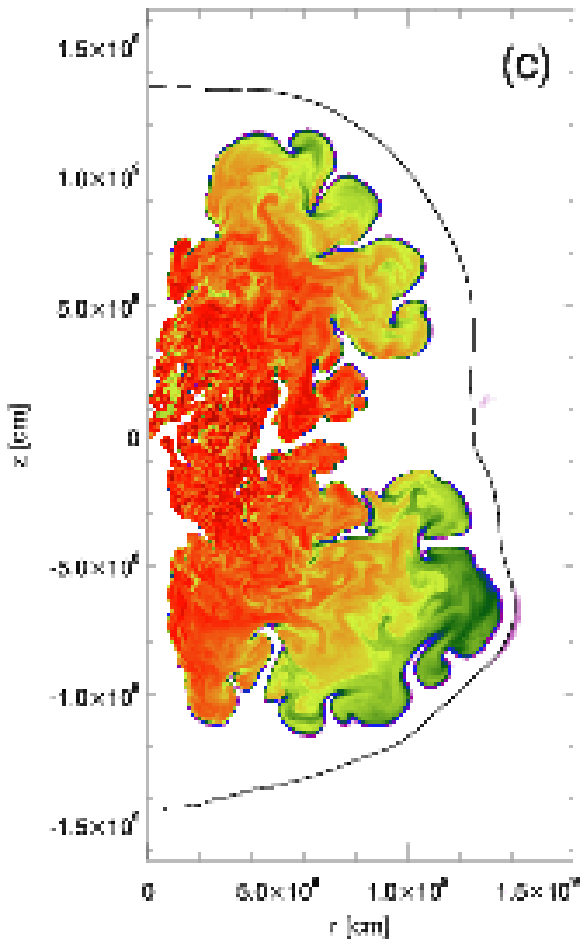}%
\includegraphics[height=5.5cm,clip=true]{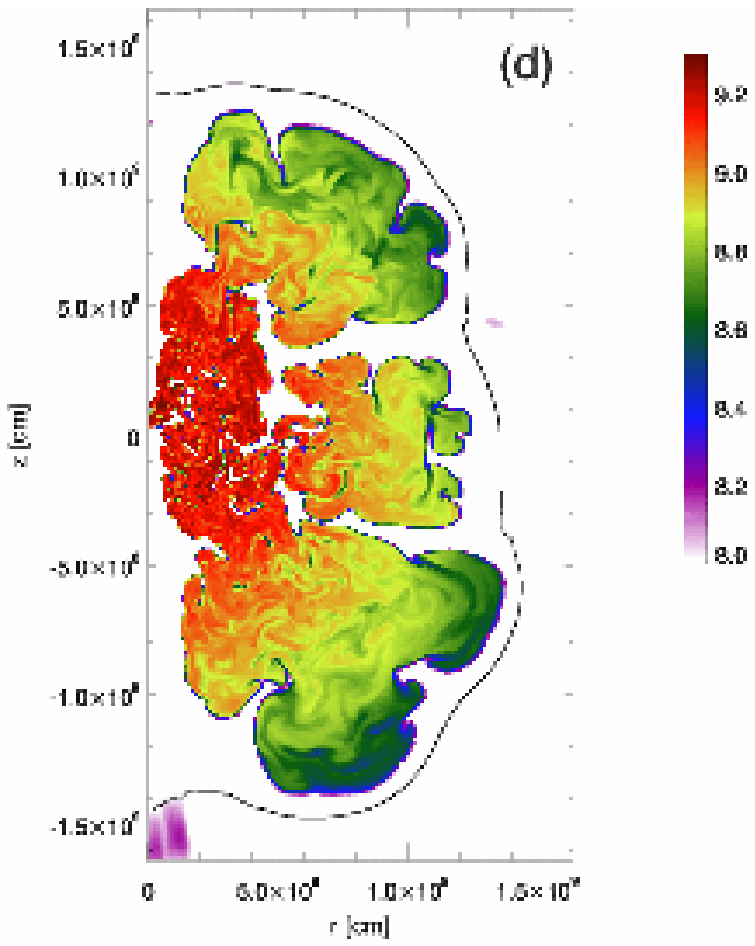}
\end{center}
\caption{Select centrally ignited benchmark supernova models from
  sequence n11d2r10t15. Panels (a)-(d) show the temperature in
  logarithmic scale at the final time ($t=2.5$~s) in models
  n11d2r10t15a, n11d2r10t15b, n11d2r10t15c, and n11d2r10t15d,
  respectively. The density isocontour at $\rho = 1\times 10^4~\gcc$
  is shown as the black solid line. Although the details of ejecta
  morphology vary strongly between models, all models produce large
  bubbles of burned material gently deforming the outer stellar
  layers. The presence of a symmetry axis in the simulation does not
  appear to bias the calculations and no asymmetry between the two
  hemispheres is observed.}
\label{f:central_morph}
\end{figure*}

As an example we compare select models from a single
sequence. Figure~\ref{f:central_morph}
%
%
%
%
%
%
shows the ejecta temperature distribution at the final time in the
sequence n11d2r10t15. The isocontour of gas density of $1\times
10^4~\gcc$ is shown with the black line and can be identified with the
stellar surface. Several comments can be made following inspection of
Fig.~\ref{f:central_morph}.

It is encouraging to notice that the models show no axis-related bias,
a numerical artifact that frequently pollutes axisymmetric
hydrodynamic calculations. In particular, the structures developed
near the symmetry axis ($r=0$~cm) in models n11d2r10t15a
(Fig~\ref{f:central_morph}(a)) and n11d2r10t15b
(Fig~\ref{f:central_morph}(b)) are markedly different. There is also
no visible difference in the amount and quality of the structure
developing in the regions above ($z > 0$~cm) and below ($z < 0$~cm)
the equator. Some models do develop structures near the equator
(see,e.g., Fig~\ref{f:central_morph}(d)), but some others do not (see,
e.g., Fig~\ref{f:central_morph}(c)). This allows us to conclude that
possible defects due to the geometry representation do not affect
our calculations in any significant way.

In all models considered here large scale structures (bubbles) several
tens degrees in size dominate in the outer parts of the ejecta. In
some models perhaps no more than 2 (Fig~\ref{f:central_morph}(c))
while in some others perhaps as many as 3
(Fig~\ref{f:central_morph}(d)) such large and distinct structures
survive turbulent burning. These bubbles push ahead unburned material
causing relatively mild deformation of the ejecta outer layers as
indicated by shape of the density isocontour.

\begin{figure*}[ht]
\begin{center}
\includegraphics[height=5.5cm,clip=true]{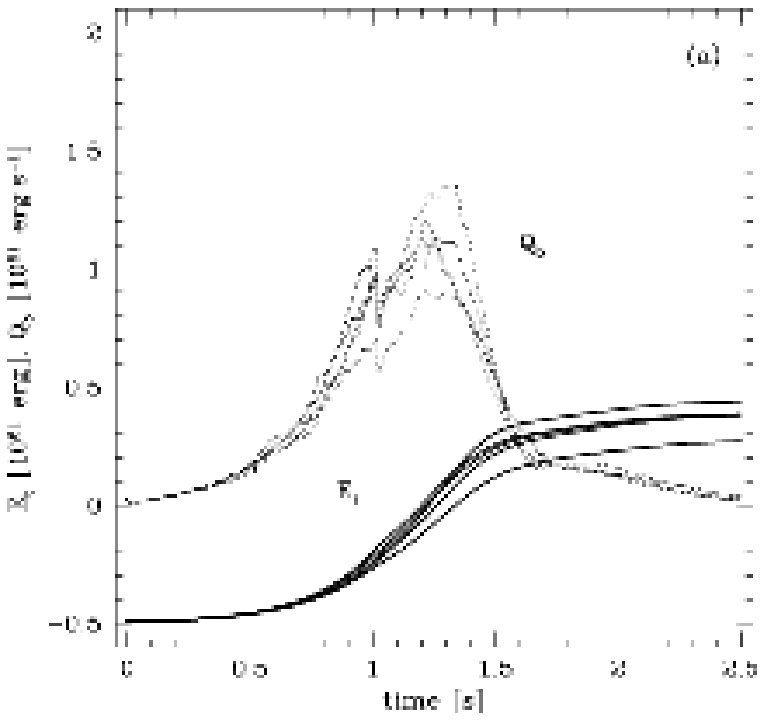}%
\includegraphics[height=5.5cm,clip=true]{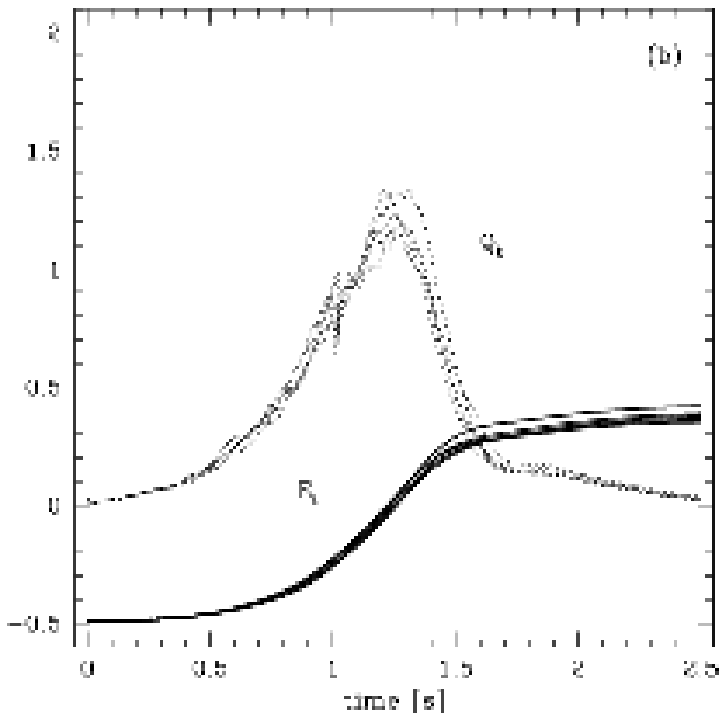}%
\includegraphics[height=5.5cm,clip=true]{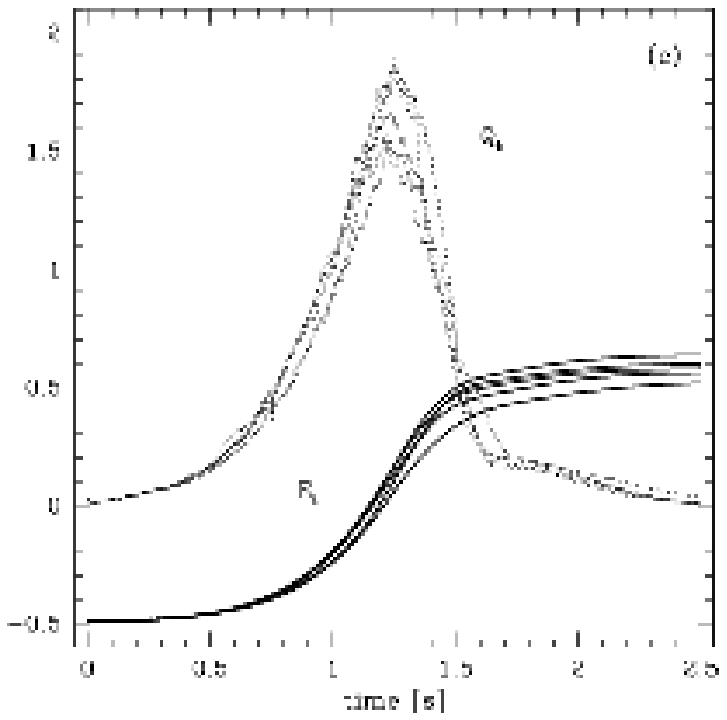}
\end{center}
\caption{Evolution of the total (explosion) energies and burning rates
 (solid and dashed lines, respectively) in three families of centrally
 ignited benchmark supernova models from sequence n11d2. (a) models
 n11d2r05t10; (b) models n11d2r10t10; (c) models n11d2r10t15. Abrupt
 changes in the energy generation rate visible at $t=1$~s (models
 n11d2r05t10 and n11d2r10t10) and $t=0.75$~s (models n11d2r05t15) are
 caused by mesh derefinement.}
\label{f:central_integrals}
\end{figure*}

As discussed below, the explosion energies also appear sensitive to
small perturbations. In the case of the n11d2r10t15 models, our
limited sample shows the total variation in explosion energy $\approx
0.1$~foe (1 foe = $1\times 10^{51}$~erg) or $\approx 10$\% in the
released energy (see Fig.~\ref{f:central_integrals}(c)). This may
indicate that small, naturally occurring variations in the internal
structure of progenitors (expected to arise from the convective flows
developing in their cores prior to runaway) may contribute to the
intrinsic diversity of SN~Ia. Addressing this interesting possibility
requires more careful study, preferably using realistic
multi-dimensional progenitor models.
\paragraph{Sensitivity to numerical resolution}
From the modeler's point of view, several simulation parameters may
potentially affect the results and therefore need to be controlled.
Given the high degree of complexity and highly non-linear character of
our application, it is natural to expect that the model results will
depend upon the numerical resolution. Convergence to the true
3-dimensional solution is not expected to occur in two dimensions due
to, for example, differences between the physics of two- versus
three-dimensional supernova turbulence
\citep{khokhlov95,schmidt+05_sgs} and Rayleigh-Taylor instability
\citep{kane+00,chertkov03}.  Nevertheless, it is still important to examine
the sensitivity of our axisymmetric model predictions to numerical
resolution.

In adaptive mesh simulations, the computational mesh is not
necessarily a well-defined entity as the numerical discretization
depends on the solution and is usually highly variable both in time
and space. The simple procedure of doubling the grid resolution does
not have its usual interpretation. Unlike uniform grid simulations,
adaptive mesh refinement computations admit additional error into the
solution by not resolving smooth or otherwise dynamically
insignificant parts of the flow field. In addition, some errors, such
as flow perturbations arising at the fine-coarse mesh interfaces, are
unique to AMR discretization and not easy to characterize
\citep{quirk91,weirs+05,pantano+06}.

As our earlier investigations have demonstrated, the morphology on
small scales of the exploding models appears very sensitive to slight
perturbations in the initial conditions, and might be useful only for
making qualitative statements. A more quantitative comparison of the
different models can be made using integral quantities. For example,
variations in the final explosion energy are of great interest from
the observational point of view, and several possible natural causes
for such variations have been proposed (i.e.\ differences in the
chemical composition and/or the rotation of the progenitor).

Here we discuss the role of mesh adaption on the evolution of the
total energies, flame surface areas and the burning rates. Since as we
mentioned before the evolution is rather sensitive to small
perturbations, we chose to compare different families of models rather
than individual family members. Figure~\ref{f:central_integrals}
%
%
%
%
%
%
shows the evolution of the integral quantities in three families of
the n11d2 sequence of benchmark models: n11d2r05t10 (left panel),
n11d2r10t10 (middle panel), and n11d2r10t15 (right panel). The
individual members of each family were obtained using slightly
different random velocity perturbations. Compared to the n11d2r10t10
models, in models n11d2r05t10 the innermost region of enhanced
resolution ($\Delta x = 2$~km) was limited to $500$~km in radius (as
compared to $1,000$~km). The resolution in this region was decreased
by a factor of 2 at $0.5$~s and by another factor of 2 at
$t=1.0$~s. In the n11d2r10t15 family of models, this innermost region
was derefined in two similar stages with the resolution ultimately
decreased to its nominal value at $t=1.5$~s.

Several observations can be made following analysis of the the n11d2
sequence.

All n11d2 models produce explosions. The explosion energies are rather
low, $\approx 0.4-0.5$~foe.  In all cases the nuclear burning is most
intense around $\approx 1.2$~s after the ignition. That period of
rapid energy release lasts for about $0.5$~s with the densities
dropping rapidly to $\approx 1-2\times 10^7~\gcc$ by $t=1.5$~s. This
is followed by a period of low, almost constant energy release during
which in only exceptional cases a slight increase in energy generation
was observed. At that time ($t\approx 1.7$~s), the burning takes place
at densities $< 1\times 10^7~\gcc$ and changes its character from
active turbulent burning associated with vigorous creation of flame
surface to a much milder, distributed mode of burning. By $t=2$~s the
densities drop to $\approx 1\times 10^6~\gcc$ and flame quenching
results in a steady decrease in energy generation.

Models obtained with higher resolution produce more energetic
explosions. The typical explosion energies vary from $\approx
0.45$~foe for the least resolved subsequence r05t10 to $\approx
0.55$~foe for the best resolved subsequence r10t15. Higher resolution
also appears to make evolutionary trajectories more similar at early
times (the curves run more closely in subsequence models r10t10 than
in r05t10) but result in increasing diversity at late times (there are
relatively large variations in energy generation rates around
$t=1.5$~s in subsequence r10t15).

We found that not only the morphology, but also the integral
quantities are sensitive to small perturbations in the progenitor. For
example, the dispersion of explosion energies is about $0.1$~foe, even
in relatively well-resolved simulations (e.g.\ subsequence r10t15).
This may indicate that some of the observed diversity of supernovae
might be produced by the nonlinear response of the explosion process
to small variations in the initial conditions. Such variations from
one progenitor to another are expected to exist in nature especially
given the convective (turbulent) flow conditions prevailing in the
stellar cores prior to thermonuclear runaway \citep[see][and
references therein]{woosley90,hoeflich+02,kuhlen+06}. The contribution
of such a purely stochastic component to the explosion process clearly
deserves more careful study.
\subsubsection{Comparison against Garching group models}\label{s:comparison}
We found good agreement between the main characteristics of our
centrally ignited model explosions and the results of equivalent
axisymmetric calculations presented by the Garching group.  The
overall evolution of the energy generation rate in the n11d2 model
sequence is similar to that in model c3\_2d\_256 by
\cite[Fig.~3]{reinecke+02_refined} and to that obtained earlier by \cite[][;
see \cite{reinecke+99}]{niemeyer95}. The energy generation in our
model explosions displays a pronounced maximum reaching between
$\approx 1.1-1.2\times 10^{51}~\ergs$ in models n11d2r05t10 and
n11d2r10t10. This compares very favorably to the result reported by
\cite{niemeyer95}, \cite{reinecke+02_refined}, and more recently by
\cite{roepke05_homo}. The latter two studies reported peak energy
generation rates $\approx 1.2\times 10^{51}~\ergs$. The rates obtained
in models n11d2r10t15 are higher by about 50\%. Such higher rates were
reported by \citep{roepke+06_diverse} for some of their centrally
ignited models, however these calculations were done in
3-dimensions. The energy generation rate in our models is initially
smaller and increases at a faster rate than in the Garching
models. The maximum generation rate is achieved at $t\approx 1.25$~s,
roughly $0.5$~s later than in the Garching models. In \FLASH\ as well
as in the Garching calculations the burning quenches $\approx 0.4$~s
after the maximum.

Our model explosions are on average slightly more energetic than those
reported in Garching studies. The typical explosion energies are
between $\approx 0.3-0.7$~foe in our models while
\citep{reinecke+02_refined} and \citep{roepke+05_distributed} reported
explosions with energies $\approx 0.35$ and $\approx 0.45$,
respectively. It is conceivable that the higher initial energy
generation rates obtained in the Garching calculations may result in
the lower explosion energies of their models since the faster initial
expansion leaves less time for the flame to develop and burn stellar
fuel.  Such discrepancies can be explained by differences in both the
adopted flame dynamics model and the approximations used to describe
the nuclear burning and not of serious concern in the context of the
following results.
\section{Detonating failed deflagrations}\label{s:dfd}
Limited by the assumption of axial symmetry, we considered off-center
bubble ignitions in which the flame initially occupies a small
spherically symmetric region(s) positioned at the symmetry axis
($r=0$~cm).  Two families of models were constructed, one with a
single ignition point and the other with two ignition points.  For the
latter, we consider only simultaneous ignitions. although in principle
the multiple ignition process could be extended in time (see, for
example,
\cite{schmidt+06_3dmpoints}).

Following our verification study, off-center explosion models were
obtained at the maximum resolution of $8$~km. In models with two
ignition points, the flame regions were initialized in different
hemispheres. Table~\ref{t:dfd_ics}
%
%
%
\begin{table}
\caption{Off-center Ignition Configurations}\label{t:dfd_ics}

\begin{tabular}{lllllll}
Model         &  $z_{b,1}$\ [km] &  $z_{b,2}$\ [km] & $R_b$\ [km]  \\
\tableline
Y12           &  12.5            &    \nodata       &   50         \\
Y25           &  25              &    \nodata       &   50         \\
Y50           &  50              &    \nodata       &   50         \\
Y100          & 100              &    \nodata       &   50         \\
Y70YM25       &  70              &     -25          &   35         \\
Y100YM25      & 100              &     -25          &   50         \\
Y75YM50       &  75              &     -50          &   50         \\

\end{tabular}

\end{table}
%
%
%
summarizes the parameters describing the initial flame and mesh
configurations of the off-center ignition models. Here $z_{b,1}$ and
$z_{b,2}$ are the locations of the ignitions points along the symmetry
axis ($r=0$~cm), and $R_b$ is the radius of flame regions. To keep the
flame regions well-separated, the mesh resolution in the central
$300$~km region was increased by a factor of 2 for a short period of
time after ignition ($0.4$~s in models Y70YM25 and Y100YM25 and
$0.1$~s in model Y75YM50). This additional refinement was also
necessary to adequately resolve the smaller bubbles ($R_b = 35$~km)
used in model Y70YM25.

The computational domain extended up to $131,072$~km and $524,288$~km
in single- and double-ignition point models, respectively. In
anticipation of an extended and asymmetric evolution at early times,
the region of adaptive meshing was extended to $6,000$~km in
radius. The initial conditions did not include random velocity
perturbations. All other simulation parameters were identical to those
used in the verification study (see \S~\ref{s:central_setup}).
\subsection{Explosion phase}\label{s:dfd_exp}
For no other reason than convenience and limits in computing power,
early multi-dimensional investigations of white dwarf deflagrations
assumed perhaps only slightly perturbed but otherwise spherically
symmetric ignition conditions
\citep{mueller+82_2ddef,livne93_2ddef,arnett+94_2ddef,khokhlov95,khokhlov00,reinecke+02_refined,gamezo+03_3ddef}.
Such a choice is not necessarily the most natural one given the white
dwarf core is believed to be convective prior to runaway
\citep{woosley90,garcia-senz+95,woosley+04}, an expectation supported by
recent multi-dimensional hydrodynamic investigations
\citep{hoeflich+02,kuhlen+06}. This led several groups to consider
progressively more complex and realistic (although not necessarily
correct!) initial flame configurations
\citep{niemeyer+96,garcia-senz+05_3dmpoints,roepke+06_3dmpoints,schmidt+06_3dmpoints}.

Here we adopt a similar approach, but not being discouraged by the
failure of an initial deflagration to produce a supernova, we continue
our investigations through the following stages of evolution. Our
preliminary investigations (PCL) indicated that the energy released in
the deflagration may be used to compress the stellar surface layers
thereby forming seed points for detonations. We were unable, however,
to study that process in detail at that time, and only speculated
about the possibility.  Here, we revisit our original idea of a
deflagration to detonation transition following a slightly off-center
ignition.
\subsubsection{Failed deflagration phase}
In all models ignited off-center, the evolution initially proceeds in
a way much similar to that described in PCL. Owing to strong buoyancy
and the relative slowness of laminar burning, the whole burning region
is quickly expelled from the core \citep{garcia-senz+95,woosley+04}
consuming only a small amount of fuel on its way to the stellar
surface. We describe the early evolution of burning regions in the set
of single ignition models in terms of mean velocities, the position of
centroids, and the effective radii of burned matter.  The effective
radii correspond to spheres of the same surface area as the burned
region.

\begin{figure*}[ht]
\begin{center}
\includegraphics[width=0.45\textwidth,clip=true]{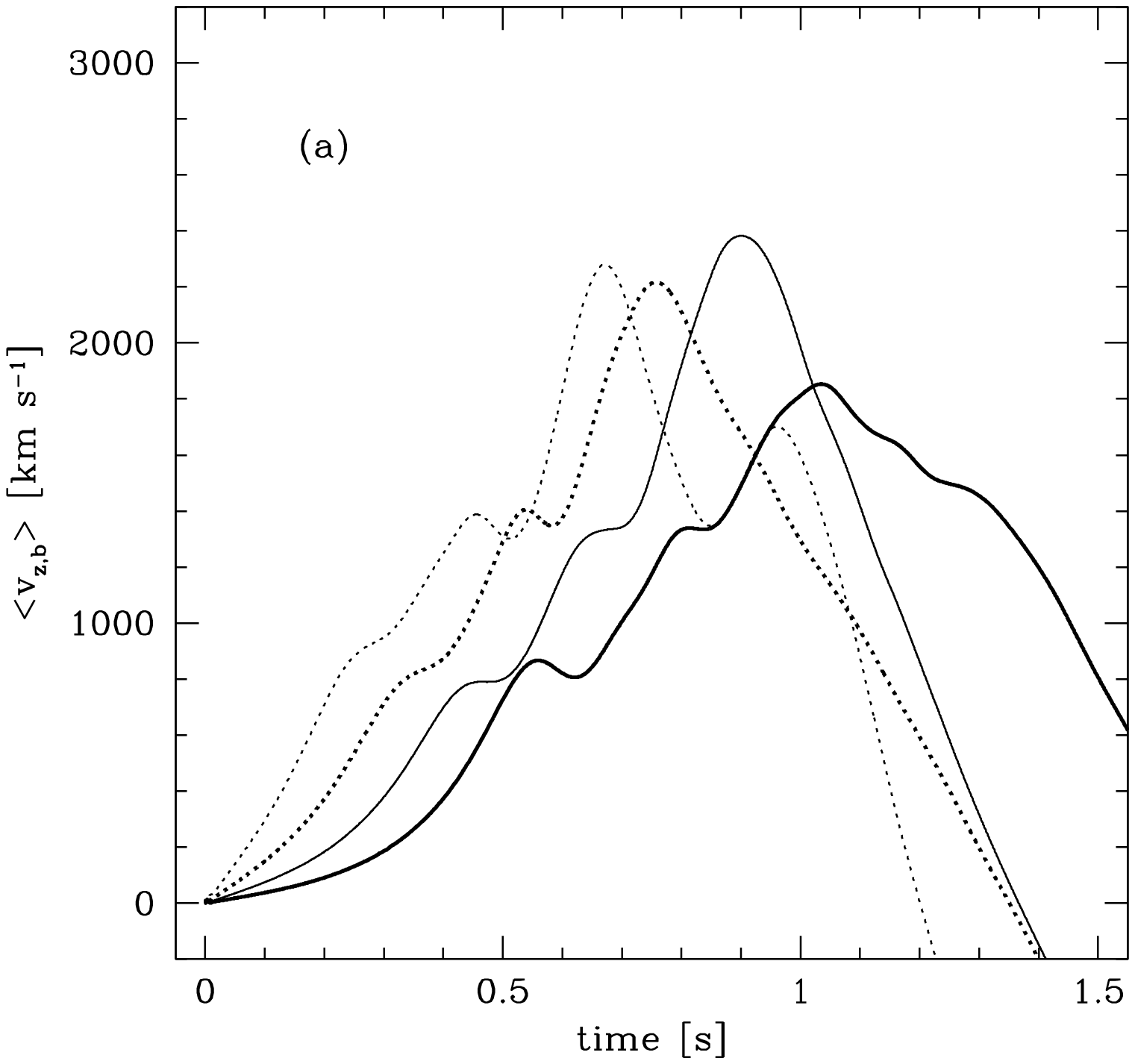}%
\includegraphics[width=0.45\textwidth,clip=true]{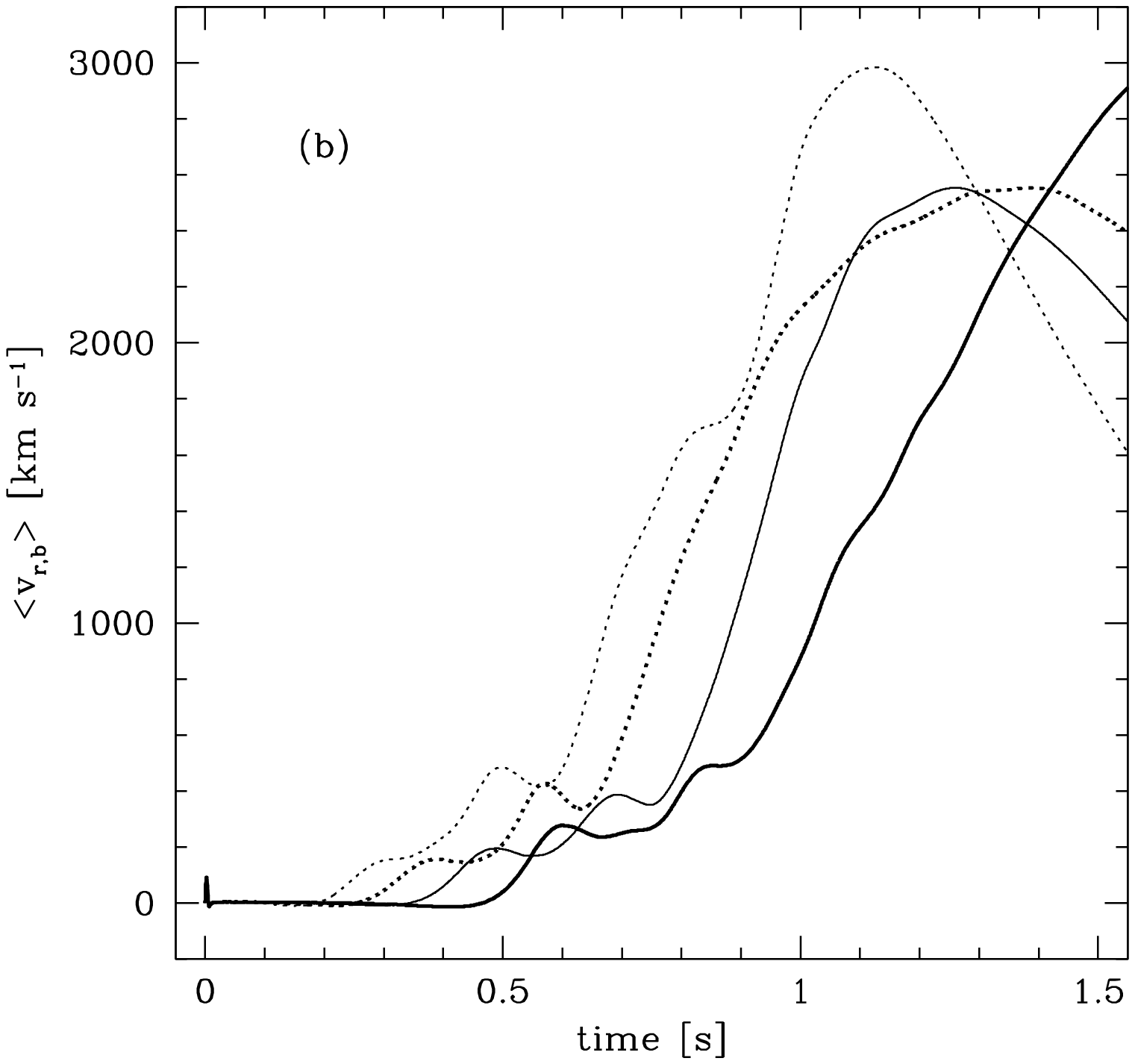}
\includegraphics[width=0.45\textwidth,clip=true]{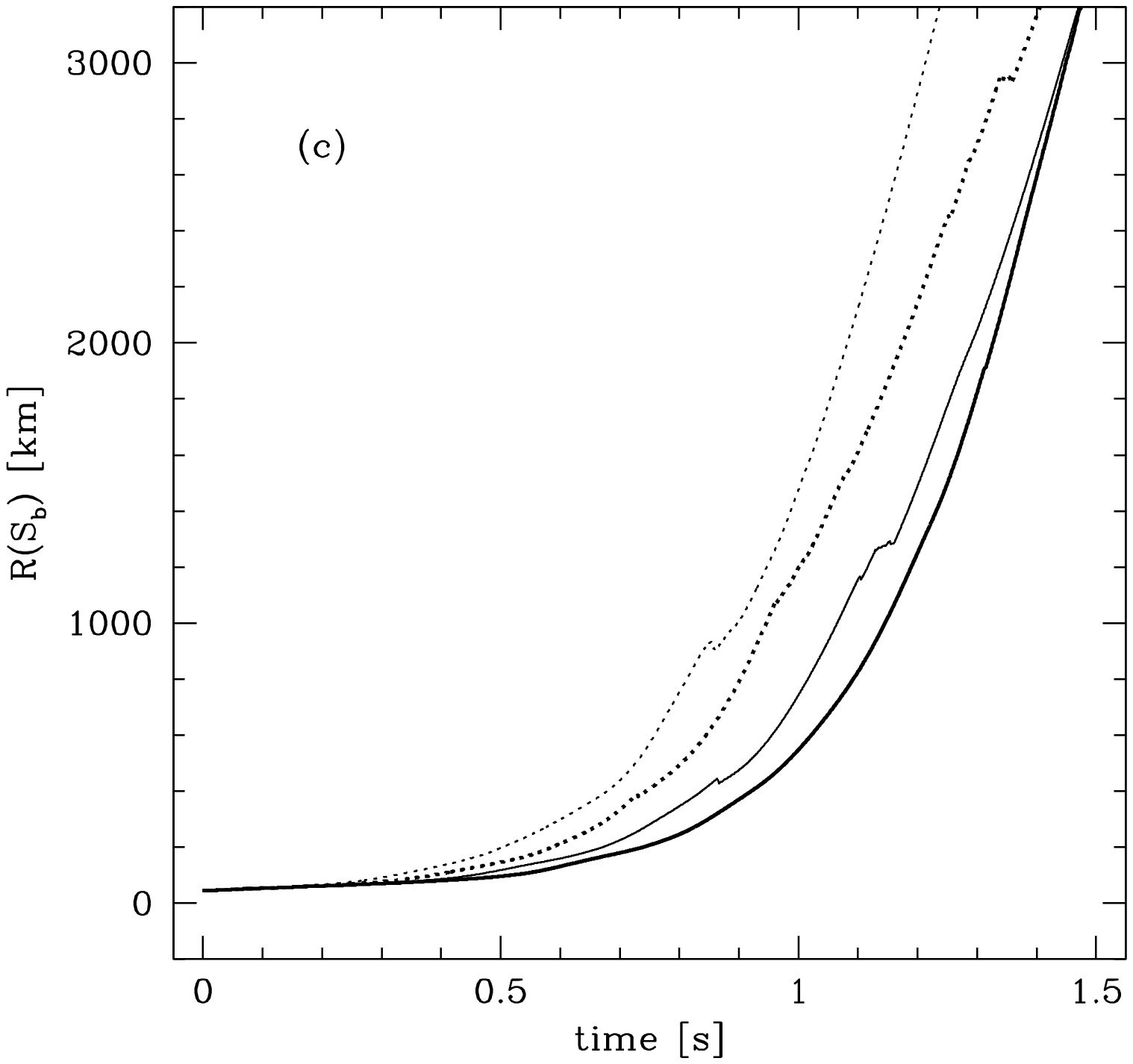}%
\includegraphics[width=0.45\textwidth,clip=true]{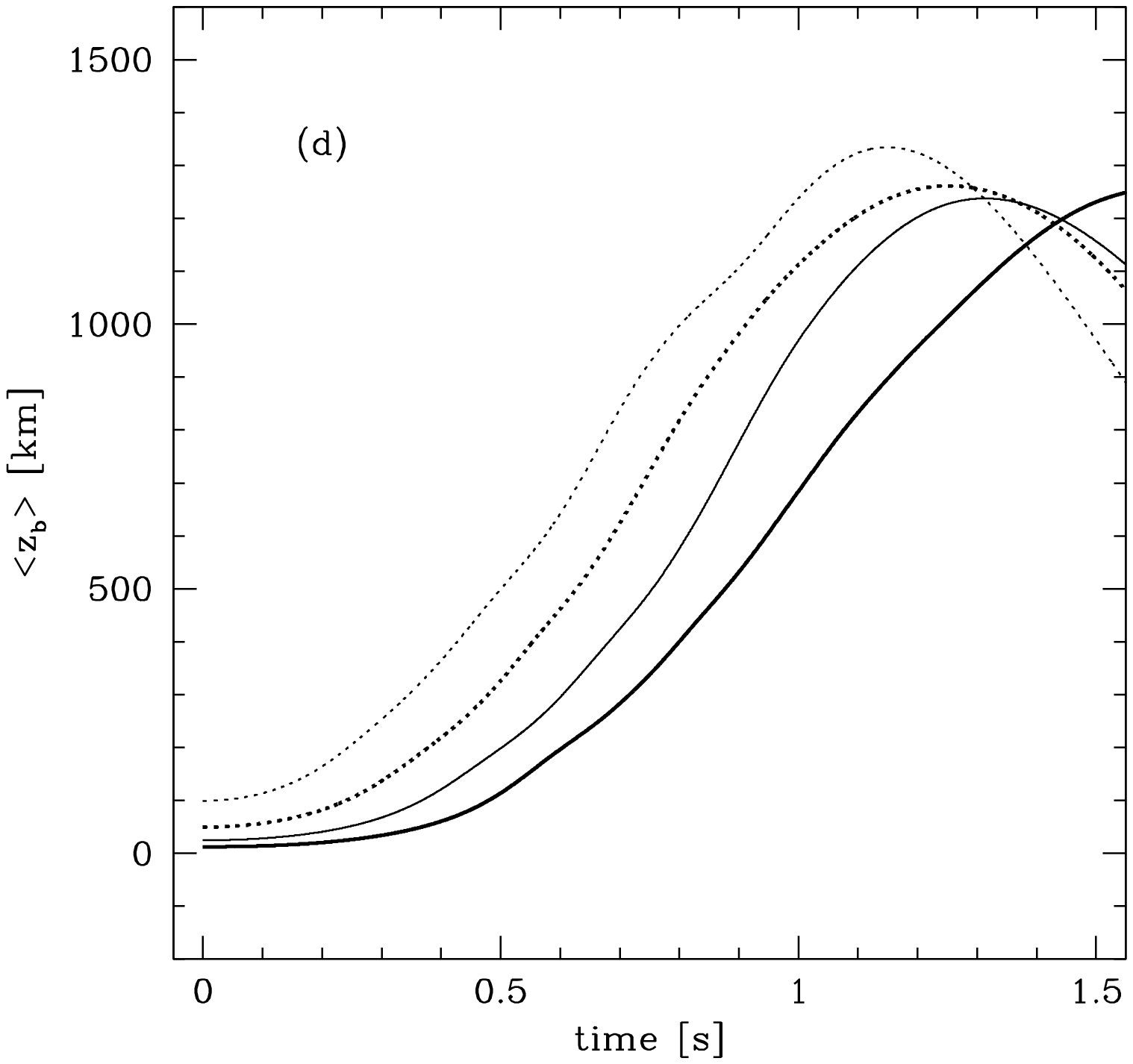}
\end{center}
\caption{Kinematics and growth of the burned region in a set of single
 ignition off-center deflagrations. (a) mean vertical velocity; (b)
 mean radial velocity; (c) equivalent radius; (d) vertical centroid
 position. Data for models Y12, Y25, Y50, and Y100 are shown with
 thick solid, thin solid, thick dotted and thin dotted lines,
 respectively.}
\label{f:bubble_kinematics}
\end{figure*}

From their original positions, the initially motionless bubbles are
driven by buoyancy and develop primarily vertical velocities, as can
be seen by comparing the mean vertical
(Fig.~\ref{f:bubble_kinematics}(a))
%
%
%
%
%
%
and radial (Fig.~\ref{f:bubble_kinematics}(b)) velocities. The initial
acceleration is stronger in cases where the bubble is placed farther
away from the stellar center (models Y50 and Y100, thick and thin
dashed lines in Fig.~\ref{f:bubble_kinematics}) and weaker when the
bubble originates deeper in the core (models Y12 and Y25, thick and
thin solid lines in Fig.~\ref{f:bubble_kinematics}). Although the mean
motion of the burned region is away from the stellar center, the rich
and complex flow field includes downflows and outflows developing
inside the region that lead to intermittent (apparent) deceleration
and acceleration (seen as mild wiggles superimposed on the velocity
curves). Each such wiggle in velocity is associated with the
destruction of the current generation of Rayleigh-Taylor bubbles and
the formation of the next. Our data indicates that perhaps two
generations of such bubbles are created during the deflagration
phase. 

\begin{figure}[ht]
\begin{center}
\includegraphics[width=0.45\textwidth,clip=true]{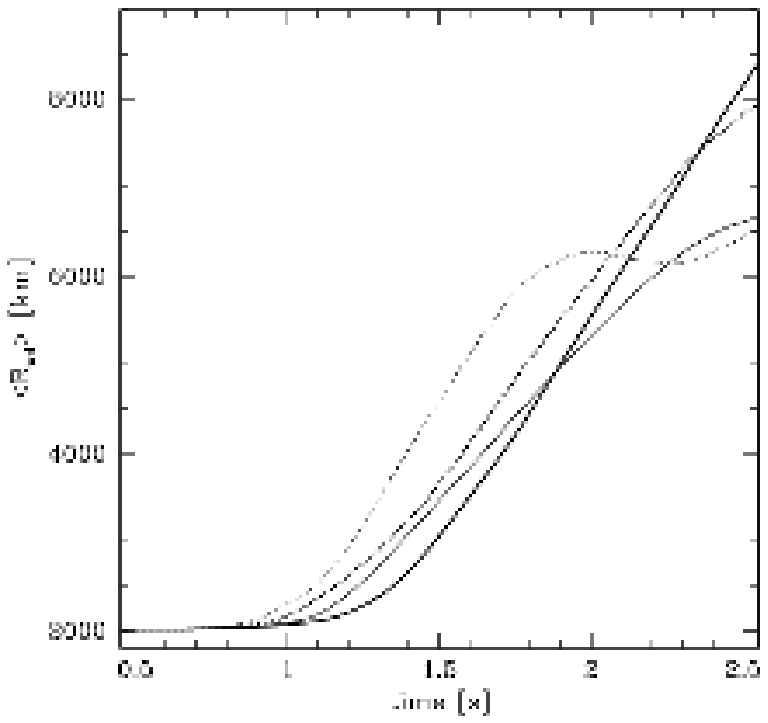}
\end{center}
\caption{Evolution of the effective stellar radius (defined as the radius of a
 sphere with the same volume occupied by matter with density $>
 1\times 10^4~\gcc$) in the single ignition off-center
 deflagrations. Data for models Y12, Y25, Y50, and Y100 are shown with
 thick solid, thin solid, thick dotted and thin dotted lines,
 respectively.}
\label{f:off_stellar_radius_single}
\end{figure}

A sudden drop in vertical velocity and a rapid increase of lateral
expansion marks the moment of bubble breakout. This phase is not
well-defined but occurs roughly at $\approx 0.7$~s in model Y100 and
not until $t\approx 1$~s in model Y12. The effective radius of the
burned regions (defined as a radius of a sphere with surface area
equal to the flame surface area) at breakout is very similar between
the models, $\approx 400-450$~km
(Fig.~\ref{f:bubble_kinematics}(c)). However, the centroid of the
burned region is located much farther out in model Y100 than in model
Y12. That is understandable considering there is more time for the RT
instability to develop structure and for the region to grow laterally
when the ignition takes place closer to the stellar center. This also
has profound consequences for the evolution of the star. A longer
deflagration phase allows for more burning, causes more matter being
lifted from the stellar core, and eventually results in stronger
stellar expansion during the early ($t < 2$~s) post-breakout evolution
(Fig.~\ref{f:off_stellar_radius_single}.
%
%
%
%
%
%
In the case of double point ignitions, the early evolution proceeds in
a very similar way to that of single ignitions with a proportionally
increase in the amount of burned material and stronger stellar
expansion.
\subsubsection{Detonation phase}
One of the motivating factors behind extending our study to double
ignition point scenarios was to examine whether a detonation can be
formed when the maximal (and perhaps even boosted by flawed numerics)
focusing offered by the symmetry axis is not present. Although we
observed detonations forming in all off-center ignited models, only in
one double ignition model, Y100YM25, does the detonation form near the
equatorial plane. In the remaining two double ignition models,
detonations eventually emerge near the symmetry axis. Although both
models eventually detonate, we cannot consider them as examples of
successful asymmetric DFDs. Nevertheless, both of them provide
additional evidence for shock to detonation transition. In what
follows we will first overview the formation of detonations in single
ignition models. Then we will discuss the flow dynamics leading to
detonation in model Y100YM25. We will conclude by presenting the
ejecta morphology soon after the shock breakout along stellar surface
is complete.

\begin{figure*}[ht]
\begin{center}
\includegraphics[width=0.30\textwidth,clip=true]{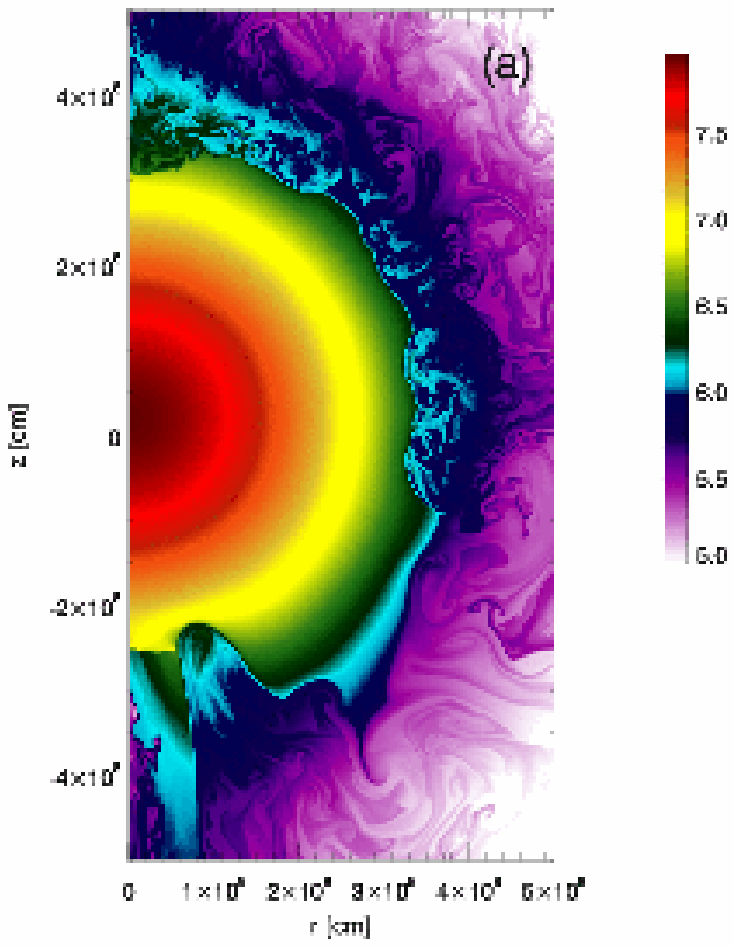}%
\includegraphics[width=0.30\textwidth,clip=true]{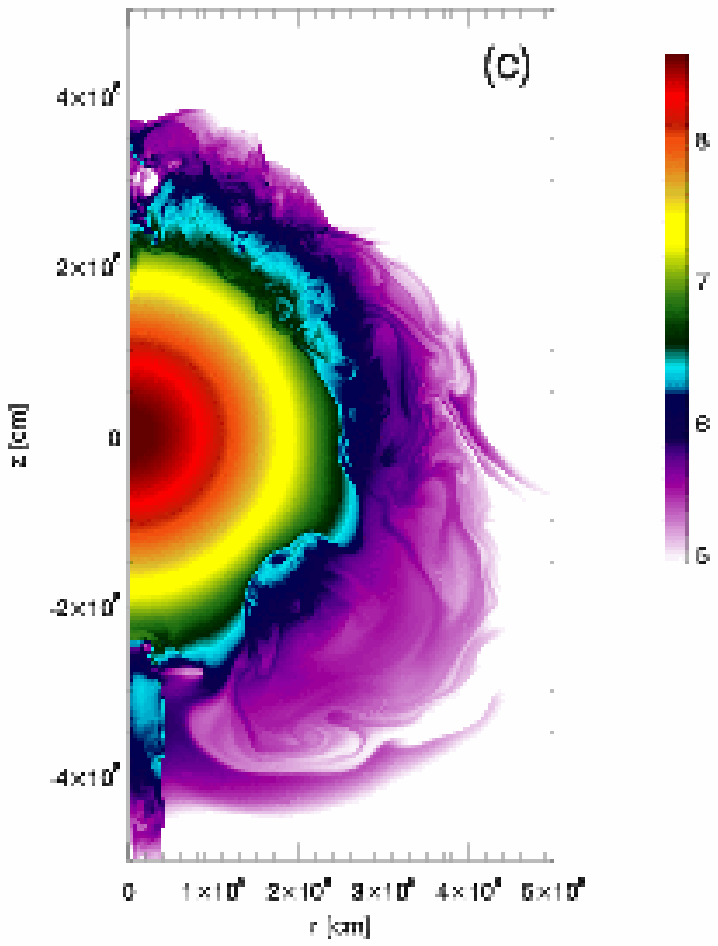}%
\includegraphics[width=0.30\textwidth,clip=true]{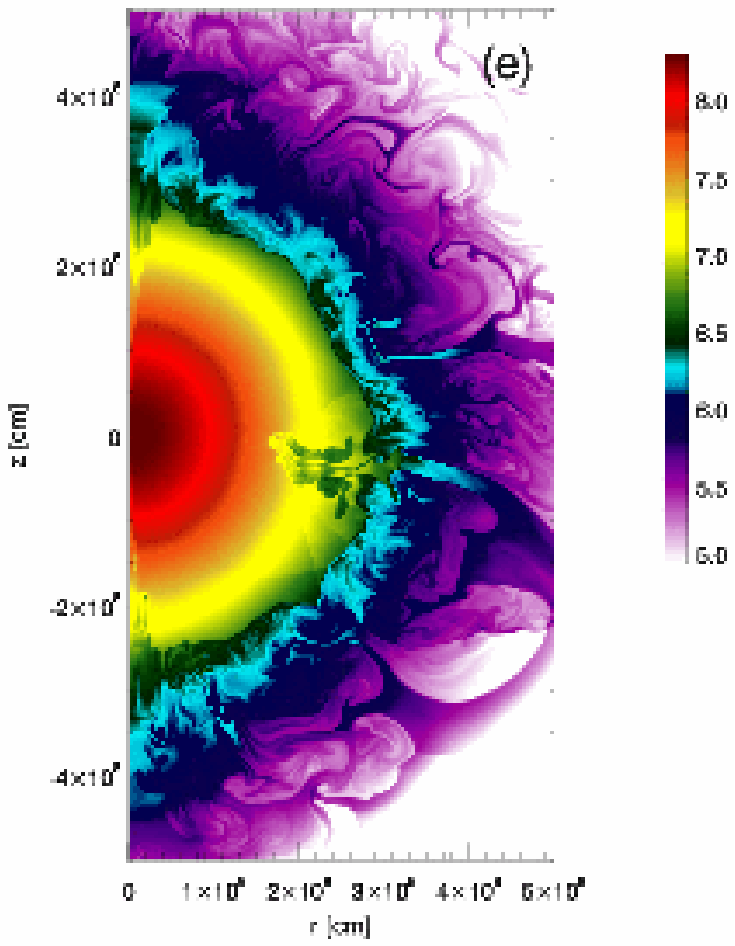}
\includegraphics[width=0.30\textwidth,clip=true]{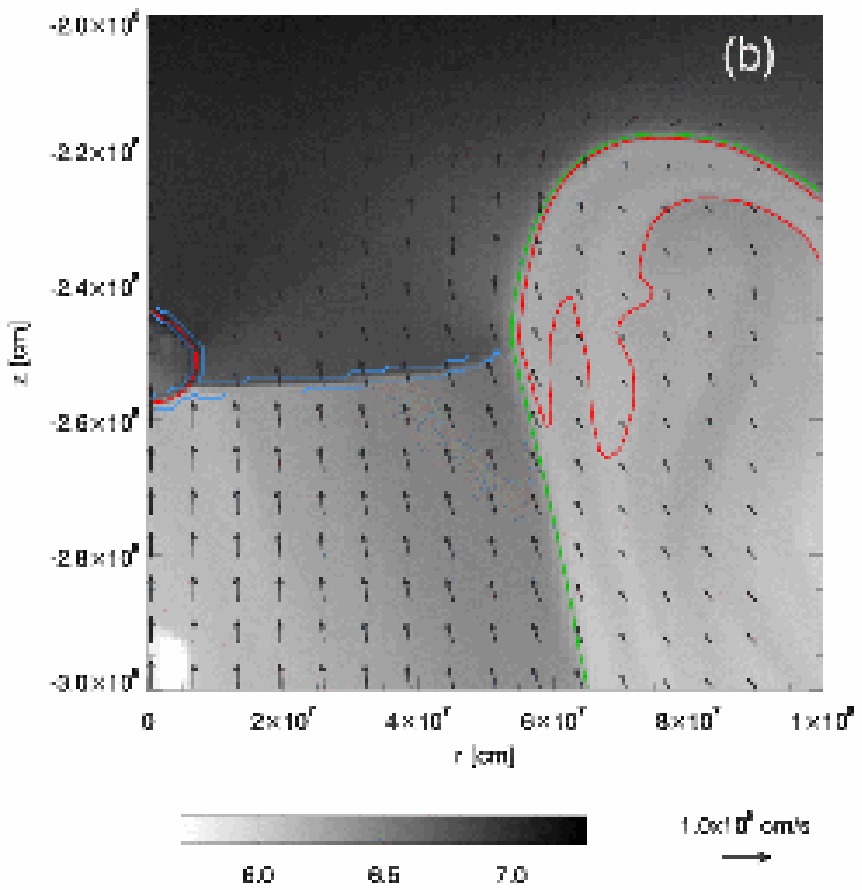}%
\includegraphics[width=0.30\textwidth,clip=true]{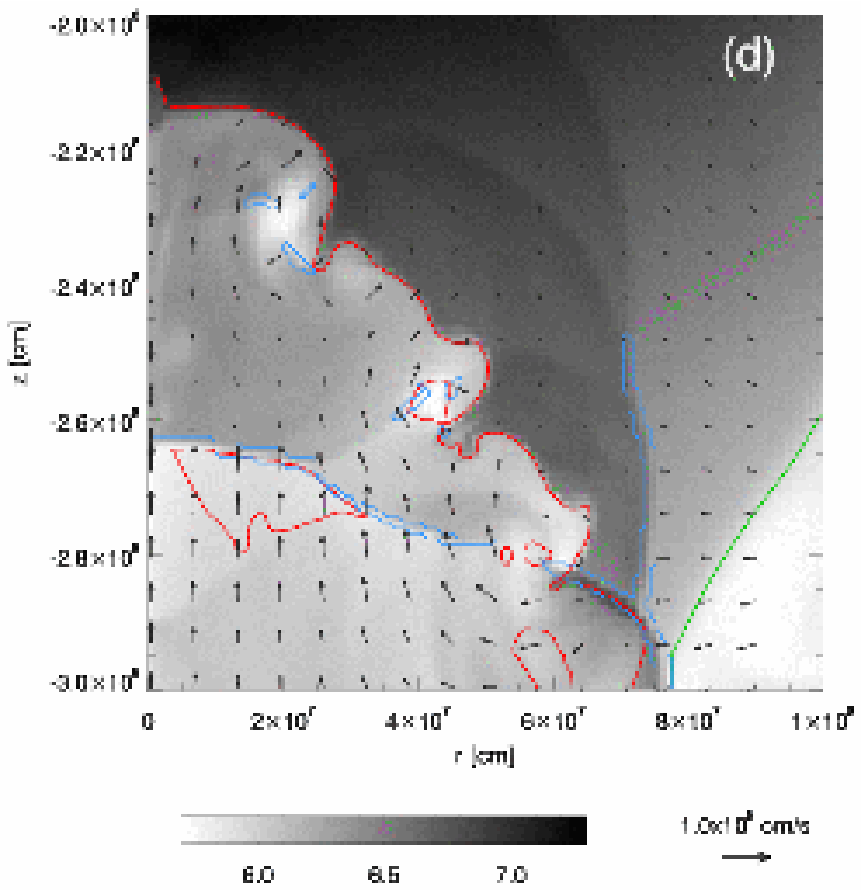}%
\includegraphics[width=0.30\textwidth,clip=true]{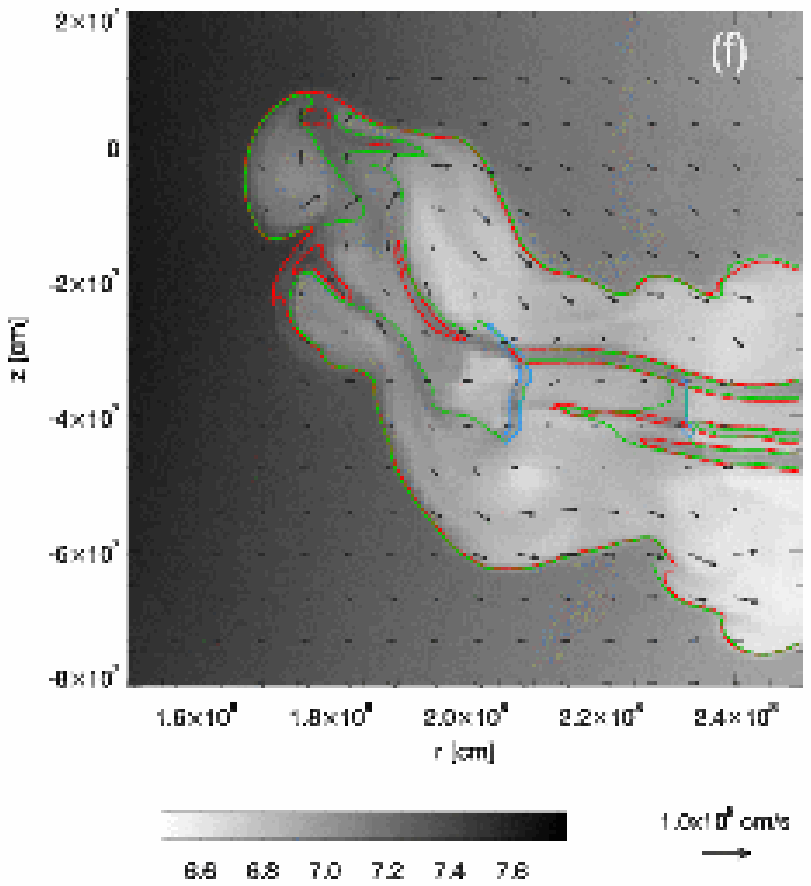}
\end{center}
\caption{Post-deflagration progenitor structure around the time of
  transition to detonation. Shown is density in log scale.  (left):
  model Y12, (a) $t=3.575$~s, (b) $t=3.6$~s; (middle): model Y100, (c)
  $t=2.0$~s, (d) $t=2.125$~s; (right): model Y100YM25, (e)
  $t=1.776$~s, (f) $t=1.8$~s.  Several structures are shown with
  isocontours in the lower row: $T = 1\times 10^9$~K (red), flame front
  (green), shocks (blue).}
\label{f:dfd1_ocd}
\end{figure*}

The progenitor structure around the time when a detonation forms is
shown in Fig.~\ref{f:dfd1_ocd}
%
%
%
%
%
%
for models Y12, Y100, and Y100YM25. In all models the bulk of
progenitor has retained its original spherical characters. We do not
find any substantial large-scale deformations, except that the stellar
cores in single-ignition models are slightly ellipsoidal in shape with
axis ratio $1.2-1.3$. Stellar expansion has decreased the core density
to $\approx 9\times 10^7~\gcc$ in model Y12 and $\approx 4\times
10^8~\gcc$ Y100. This is consistent with our expectation that lower
central densities are to be found in models that experienced more
energetic deflagrations.

The stellar core is surrounded by an extended strongly turbulent
atmosphere. Comparing three models shown in the upper row in
Fig.~\ref{f:dfd1_ocd}, the atmosphere appears better developed (more
extended and turbulent) in models that release more energy in the
deflagration. This atmosphere formed following the breakout of
deflagration products through the stellar surface, at which time the
ashes accelerated unburned surface layers both radially and
laterally. This circular wave carried both fuel and products of the
deflagration along the surface of the star. The following evolution
depends on whether there was one or more ignition points.

If the single ignition case, the surface wave eventually completely
engulfs the progenitor and collides with itself in a region located
opposite breakout. A conical shock wave forms in the process that
thermalizes the kinetic energy of the incoming flow. This shock can be
seen as a vertical structure near $r=0$~cm extending down from
$(r,z)\approx (8\times 10^7,-3.5\times 10^8)$~cm in model Y12
(Fig.~\ref{f:dfd1_ocd}(a)) and $(r,z)\approx (4\times 10^7,-2.5\times
10^8)$~cm in model Y100 (Fig.~\ref{f:dfd1_ocd}(c)).

In the multi-point ignition case, there will presumably be several
breakout points and related surface waves that will be colliding with
one another. Therefore, it is conceivable that several shock-dominated
regions might be formed. Some of those shocks might be weaker and
others stronger than the ones found here. The number of possible
scenarios and outcomes is certainly much larger than represented in a
limited sample of the initial configurations considered in this
study. Nevertheless, we expect that our models capture the essential
features of the evolution. In the case of model Y100YM25, for example,
the material of two waves collides near the equatorial plane forming a
jet-like radial inflow and outflow near $(r,z)\approx (2.5\times
10^8,-5\times 10^7)$~cm (Fig.~\ref{f:dfd1_ocd}(e)). This is
essentially the same flow configuration we found in the single point
ignition models.

In what follows, we first focus our discussion on details of the
transition to detonation process in three selected models. Then we
characterize the evolution of the exploding star during the passage of
the detonation wave.
\paragraph{Shock-to-detonation Transition}
In all cases considered in our study, we found explosions following a
shock-to-detonation transition, SDT \citep[see, e.g.,][and references
therein]{bdzil+92,sharpe02}. Although regions forming detonations
differ greatly in morphology, the common ingredients of the process
include the presence of a strong acoustic wave, dense fuel, and a
prolonged compression of the region. This is illustrated in the bottom
row of Fig.~\ref{f:dfd1_ocd} which shows the density distribution and
major flow structures involved in a transition to detonation process.

Model Y12 shares the initial conditions with the original PCL study.
In the Y12 model, we found no sign of a possible transition to
detonation for $t < 3.5$~s (Fig.~\ref{f:dfd1_ocd}(a)) in contrast to
the PCL study, in which a detonation wave formed at $t\approx 1.9$~s
This difference in timing is due solely to the incorrect energetics of
approximate deflagration burner used in the PCL calculations. The
overestimate of the energy release and hence buoyancy by a factor of
$\approx 3$ in the PCL model resulted in a much shorter deflagration
phase, a lower overall energy release, a more compact progenitor and
ultimately a significantly earlier formation of the detonation.  At
the same time, the less expanded progenitor allowed for the surface
wave to move at a relatively higher speed (due to a lower orbit) and
with higher post-shock temperatures, possibly enhancing the likelihood
of a transition to detonation.

Similar to PCL, we observe the formation of a conical shock in
Y12. However, the conditions inside the shocked region allow only for
residual burning that perturbs the shocked gas (the low density region
near symmetry axis at $r=0$~cm in the lower section of
Fig.~\ref{f:dfd1_ocd}(a)). The wave that transits to detonation is an
accretion shock (blue isocontour extending horizontally from
$(r,z)\approx (0,-2.55\times 10^8)$~cm in Fig.~\ref{f:dfd1_ocd}(d))
created by infalling material which is trapped horizontally by the
symmetry axis and the incoming deflagration products, and vertically
by the bulk of the stellar material and the stagnation point formed
behind the conical shock. A transition to detonation takes place at
the symmetry axis where the ram pressure of the incoming flow and the
resulting post-shock temperature are the highest. (Although the
accretion flow is predominantly along z-direction, there exists a
lateral velocity gradient, $\mathrm{d}v_r/\mathrm{d}r\approx -3$, in
the flow that focuses the flow toward the symmetry axis.)

In model Y100 (Fig.~\ref{f:dfd1_ocd}(d)), a detonation wave can be
seen as a nearly vertically propagating shock located near
$(r,z)\approx (6.5\times 10^7,-2.85\times 10^8)$~cm. Unlike in model
Y12, here the detonation is not directly associated with the symmetry
axis. The conical shock is visible as an wave originating near
$(r,z)\approx (7.5\times 10^7,-3\times 10^8)$~cm, just to the right of
the detonation region. The apparent closeness of the flame front is
coincidental and has nothing to do with origin of the detonation
wave. By the time of Fig.~\ref{f:dfd1_ocd}(d) the conical shock is
already sweeping through deflagration products, however the detonation
appears moving through a channel of shocked fuel eventually connecting
to the bulk of stellar material.

Model Y100YM25 displays by far the most complex flow structure in the
region where a transition to detonation occurs. The configuration of
two colliding surface waves resembles that of a ``self-colliding''
surface wave found in models with a single ignition point.  Here,
however, a symmetry of the problem is broken. Not only does the
collision not take place at the symmetry axis, but the ignition points
were initially located at different distances from the core.  The
timing, energetics, and morphology of each wave were therefore
slightly different. This difference eventually leads to a shift of the
collision plane $\approx -400$~km from the equator
(Fig.~\ref{f:dfd1_ocd}(f)). Furthermore, the collision does not occur
``head-on'' but rather material from the slower wave (ignition point
located closer to the core) tends to penetrate underneath that of the
faster wave (ignition point located farther away from the core). In
the end, the whole region shows a slight tendency to roll.

Another difference from the highly symmetric single ignition models is
that the broken symmetry offers the potential for creating more than
just one shocked region. This is indeed the case in model
Y100YM25. Two shock fronts moving in the radial direction can be seen
inside the collision region: one located near $(r,z)\approx (2.3\times
10^8,-4\times 10^7)$~cm, and another near $(r,z)\approx (2.1\times
10^8,-4\times 10^7)$~cm. Both fronts are created near the collision
plane which might be understood given this is where we expect the
thermalization rate of the colliding flows being the greatest. The
former shock wave evolves into a self-sustained detonation while the
latter soon dies off. Once a detonation is formed, the wave travels
approximately along a fuel-rich channel (a flame-bounded horizontally
extending structure near $(r,z)\approx (2.5\times 10^8,-3.5\times
10^7)$~cm that connects to the bulk of the unburned stellar material.

\begin{figure}[ht]
\begin{center}
\includegraphics[width=0.45\textwidth,clip=true]{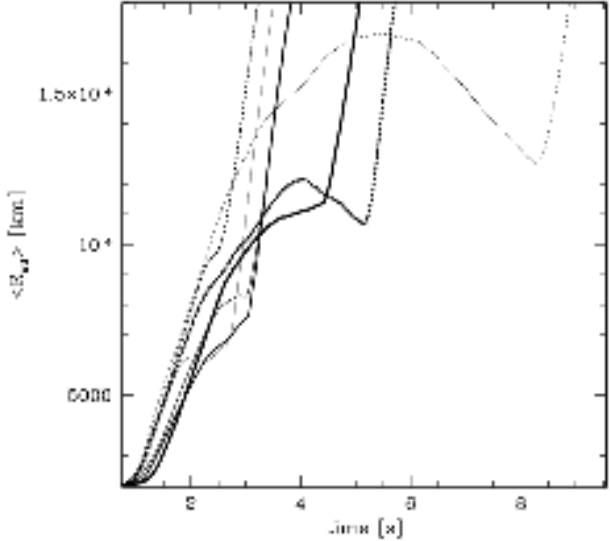}
\end{center}
\caption{Evolution of effective stellar radius (defined as radius of a
 sphere with the volume occupied by matter with density $> 1\times
 10^4~\gcc$) in model Y12 (thick solid), Y25 (medium solid), Y50
 (solid), Y100 (thin solid), Y70YM25 (thick dotted), Y100YM25 (medium
 dotted), and Y75YM50 (dotted). Evolution toward SDT proceeds through
 overall continuous expansion followed by a possible period of
 recollapse. Shock breakout is followed by a very rapid expansion of
 stellar material at approximately constant velocity.}
\label{f:off_stellar_radius}
\end{figure}

Some more details and observations can be offered regarding the SDTs
observed in our subset of models. We found that transitions to
detonation occur in gas with pre-shock densities $\approx 1-3\times
10^6~\gcc$ in models Y12 and Y100 and $> 5\times 10^6~\gcc$ in model
Y100YM25. As demonstrated by \citep{arnett+94_2dddt}, at these
densities the typical radius of region that can successfully transit
to detonation is $\sim$few kilometers. This is smaller than the
numerical resolution in our models. It is conceivable that problems
with producing SDTs in some of our double-point ignition models might
be attributed to insufficient resolution. For the same reason, the
observed SDTs may require less time to launch detonations after a
strong shock wave forms. However, preconditioning of the fuel for SDT
is certainly a temporally extended process requiring both compression
of the material and thermalization of the flow. The latter is aided by
the confinement that makes the thermalization process more
efficient. Still, large amounts of energy need to be supplied to the
region and the typical velocity jumps across shock waves are
$4,000-6,000~\kms$. This guarantees post-shock temperatures $> 1\times
10^9$~K, high enough for nuclear burning to take control of the flow
dynamics.
\paragraph{Evolution through detonation}
Transitions to detonations occur in different models at different
times and locations although, as we discussed earlier, several
necessary elements (high density fuel-rich matter, strong wave,
extended confinement of the region) are commonly
present. Figure~\ref{f:off_stellar_radius}
%
%
%
%
%
%
shows the evolution of the equivalent stellar radius (defined as the
radius of a sphere with the same volume as that occupied by matter of
density $> 1\times 10^4~\gcc$) in our sample of DFD models. This
initial steady stellar expansion is due to energy deposition by a
failed deflagration. In most cases a detonation occurs when the
progenitor either approaches or begins to recollapse. This is
understandable since at later times the energy of surface waves is
likely to quickly dissipate, thus decreasing the likelihood of strong
hydrodynamic interactions taking place (e.g.\ in multi-point
ignitions, model Y100YM25). Alternatively, the accretion flows can
develop only once expansion stops and high accretion luminosities
(required for SDT) can be expected only shortly after accretion flows
develop. In particular, in models Y70YM25 and Y75YM50 (thick dotted
and dotted line in Fig.~\ref{f:off_stellar_radius}, respectively) no
SDT occurred during the collision of surface waves. Instead, flow
perturbations accumulated in the regions near the symmetry axis,
evolved into jet-like flows and eventually triggered detonations. With
our verification study providing evidence that the evolution of
perturbation near the symmetry axis cannot be entirely trusted, we do
not consider these two models as successful DFDs produced by double
point ignitions. (One probably could still consider them members of
single point ignition families, perhaps obtained from different
initial conditions, and provided their radii prior to SDT were similar
to original single ignition models. We do not consider this inelegant
possibility any further.)

\begin{figure*}[ht]
\begin{center}
\includegraphics[height=5.5cm,clip=true]{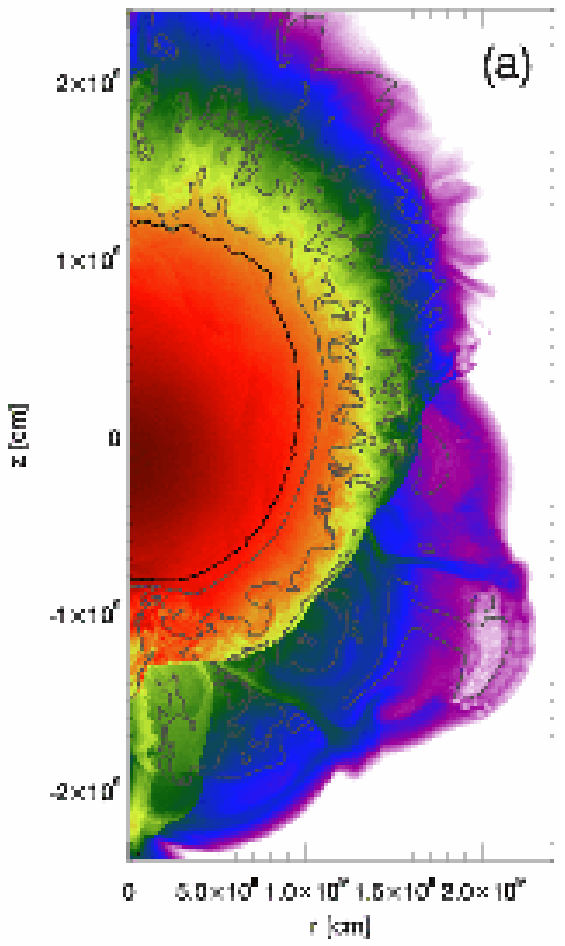}%
\includegraphics[height=5.5cm,clip=true]{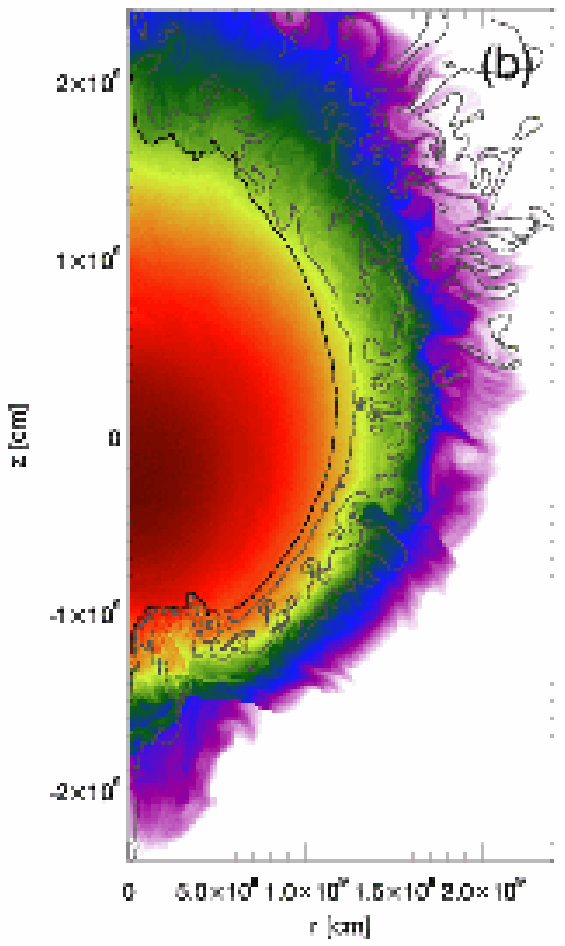}%
\includegraphics[height=5.5cm,clip=true]{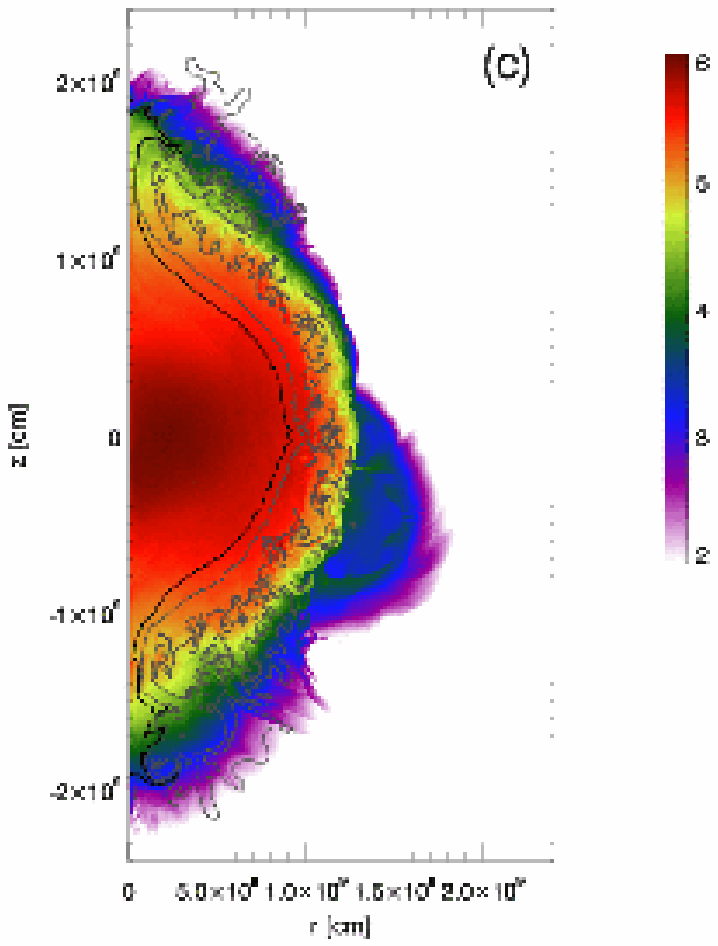}
\end{center}
\caption{Post-detonation structure of the exploding supernova at the
  time when the central density drops to $1\times 10^6~\gcc$ and
  burning effectively quenches. Panels (a)-(c) show the density in log
  scale in models Y12 ($t=4.90$~s), Y100 ($t=3.375$~s), and Y100YM25
  ($t=2.878$~s). Abundance isocontours $X(\nuc{56}{Ni})=0.95$ and
  $X(\nuc{28}{Si})=0.2$ are shown with black and gray lines,
  respectively.}
\label{f:dfd1_ocpd}
\end{figure*}

Figure~\ref{f:dfd1_ocpd}
%
%
%
%
%
%
shows the morphology of the exploding model supernovae Y12, Y100, and
Y100YM25 shortly after the central density drops below $1\times
10^6~\gcc$ and nuclear burning essentially quenches. Several important
observations can be made. All model supernova ejecta are
stratified. The ejecta are composed of a featureless core surrounded
by inhomogeneous external layers. This composite structure of the
ejecta is a direct result of the two stages of the explosion, each
involving a diametrically different dominant process. The core is the
remnant of the original progenitor which has been expanded by energy
released during the deflagration phase. the density distribution in
the core displays slight asymmetry reflecting the character of the
initial conditions (the shift in the density maximum to the lower
hemisphere present in single ignition models is absent in double
ignition model) . The outer layers may show global asymmetry,
especially in models with a single ignition point, but are always rich
in structures reflecting the turbulent nature of the deflagration and
the violent evolution of surface waves.

The density stratification of the ejecta is accompanied by
compositional stratification. This is due to the detonation wave which
synthesized iron peak elements when sweeping through the dense central
regions and intermediate group elements when encountering the less
dense outer layers.  The detonation is driven by essentially
instantaneous energy release due to carbon burning followed by
approach to NSQE (silicon-group elements) and final relaxation to NSE
(iron-group elements). With relatively crude resolution, only the
relaxation to NSE and approach to NSQE at the lowest densities can be
considered spatially resolved in our simulations
\citep[Fig.~9]{khokhlov89}. This problem, however, is not related to
the energy release and so does not influence the overall dynamics of
the detonation front. The low resolution of our models prevents us
from considering possible effects related to the curvature of the
detonation front. Once again, such effects affect the structure of the
detonation wave only on scales $\simlt 10$~km
\citep{sharpe01} and are not expected to influence the large scale
dynamics of the detonation wave.

The outermost ejecta are composed of unburned fuel mixed with the
deflagration products, most likely intermediate group elements with
locally entrained iron-rich material. Close to the core, those layers
were overrun by the detonation wave that additionally modified their
composition. Although definitely present in our calculations, we
estimate this effect to be small.  To calculate the detailed
composition of the ejecta, including the deflagration mix, requires
postprocessing nucleosynthesis \citep[see for
example][]{travaglio+06_metallicity} and is beyond the scope of this
presentation.

Detonations are known to be susceptible to transverse perturbations
and developing cellular structure
\citep[Chap. 7]{fickett+01}. Potential sources for such perturbations
are abundant in our models and include upstream flow perturbations due
to preexisting convection and turbulence (not considered here),
initial deflagration, or possible numerical oscillations of
grid-aligned shock fronts, i.e.\ odd-even decoupling
\citep{quirk94}. We found no clear evidence for cellular structure in
our calculations. One possibility is that the numerical resolution is
insufficient to resolve detonation cells of sub-km size
\citep{gamezo+99,falle00}. Also, the time available for cellular
structure is very limited as low density material rapidly expands
following the passage of the detonation wave and the burning quickly
quenches.

\begin{figure}[ht]
\begin{center}
\includegraphics[width=0.45\textwidth,clip=true]{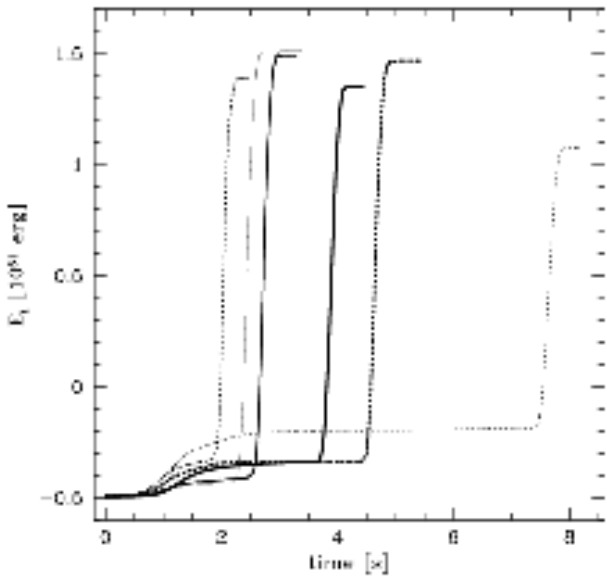}
\end{center}
\caption{Evolution of the total (explosion) energy through the end of
 detonation phase in model Y12 (thick solid), Y25 (medium solid), Y50
 (solid), Y100 (thin solid), Y70YM25 (thick dotted), Y100YM25 (medium
 dotted), and Y75YM50 (dotted). Only small amount of energy, $\approx
 0.06-0.15$~foe is released during deflagration phase. Detonation
 phase lasts $\simlt 0.4$~s.}
\label{f:off_stellar_energy}
\end{figure}

The evolution of the explosion energy (kinetic + internal + potential)
and burned mass from the moment of ignition until the shock breakout
is shown in Fig.~\ref{f:off_stellar_energy}.
%
%
%
%
%
%
With plenty of unburned fuel available to the detonation, DFD
supernovae are energetic events with typical post-detonation energies
$\approx 1.3-1.5$ foe.  The deflagration phase typically supplies only
$0.06$ to $0.15$~foe of energy in burning $< 0.1~\msun$ of stellar
fuel. The bulk of the energy is released during about $0.4$~s when the
detonation wave sweeps through the white dwarf. These findings
resemble the results presented by \cite[Fig.~3]{arnett+94_2dddt}
although the mechanism behind the transition to detonation is
different in the two models.
\subsection{Homologous expansion: final properties}\label{s:dfd_homo}
The direct comparison of hydrodynamic explosion models to observations
is accomplished through the calculation of synthetic model light
curves, spectra, and possibly also spectrum polarization. These
radiative transfer calculations take as an input the model supernova
ejecta with complete specification of the density distribution, ejecta
chemical composition, and typically make the simplifying assumption
that velocity linearly increases with distance from the ejecta
center. This last assumption is satisfied to different degrees in
various parts of the ejecta and generally does not hold true during
the early stages of supernova expansion. The reason is that linear
expansion requires establishing a fine balance of accelerations
between neighboring fluid elements in the ejecta, and this requires
time. For example, our post-detonation explosion models contain large
regions where energy of the flow is dominated by internal energy. This
indicates the potential of fluid elements performing some work,
possibly adjusting their motion relative to their neighbors. To allow
for that process to operate and establish homologous expansion, the
post-explosion needs to be continued for an extended period of time.
Detailed discussion of this process in the context of Type~Ia
supernova modeling was recently presented by
\cite{roepke05_homo}.

We obtained a complete set of homologously expanding model ejecta
using \FLASH\ and its adaptive mesh capabilities. Post-detonation
models were carefully transported to a high-resolution uniform mesh
with typical relative errors of total mass, total energy, and
abundances not exceeding $0.1$\%, $0.5$\%, $5$\%, respectively. Given
the several sources of uncertainties and variations in the original
models, this accuracy is sufficient for any practical purposes. The
interpolated models were then used to define the initial conditions in
the \FLASH\ calculations.

We used a ratio of kinetic energy to the sum of internal plus
gravitational energies to monitor the approach of ejecta to
homology. In course of several trial computations, we established that
continuing calculations for $100$~s after explosion guarantees that
the energy ratio $> 100$ anywhere in the ejecta except for the
innermost half of the central nickel-rich core (and parts of
essentially massless shocked ambient medium). This required using a
computational grid extending to $1.68\times 10^{12}$~cm. Calculations
were performed using an automatic mesh derefinement scheme that kept
the computational cost approximately constant as the ejecta
expanded. The effective mesh size varied from $8,192$ initially to
$2,048-4,096$ zones per dimension at a final time.

\begin{figure*}[ht]
\begin{center}
\includegraphics[height=5.5cm,clip=true]{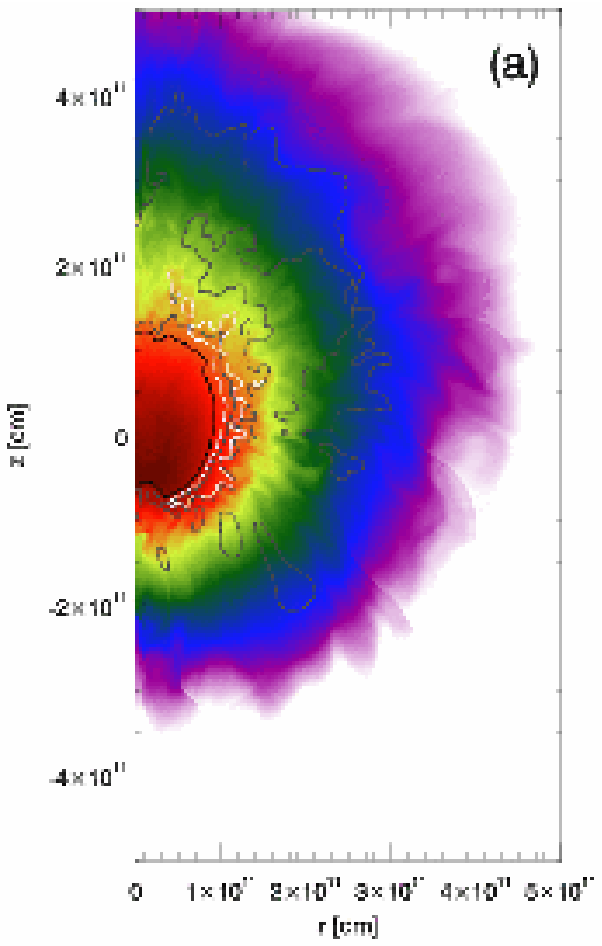}%
\includegraphics[height=5.5cm,clip=true]{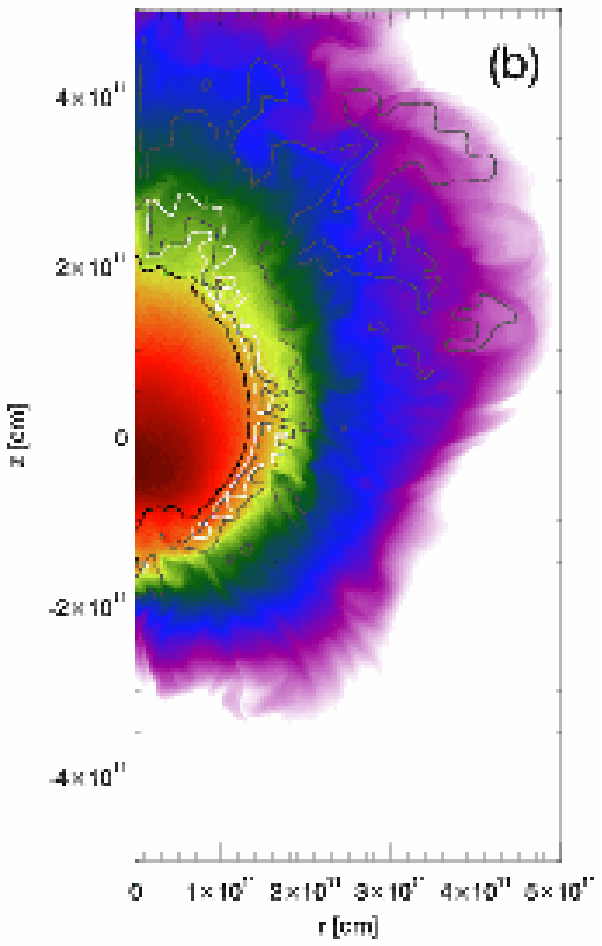}%
\includegraphics[height=5.5cm,clip=true]{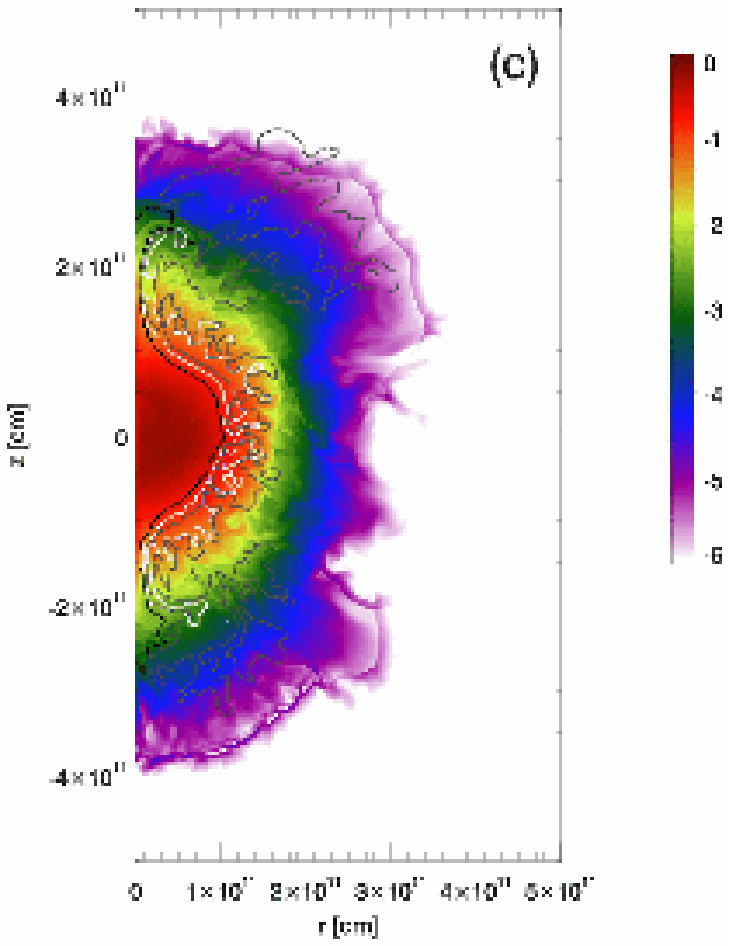}
\end{center}
\caption{Homologously expanding model ejecta $\approx 100$~s after
  explosion. Panels (a)-(c) show density in log scale in models Y12,
  Y100, and Y100YM25. Abundance isocontours $X(\nuc{56}{Ni})=0.95$,
  $X(\nuc{28}{Si})=0.2$, and $X(\nuc{40}{Ca})=0.05$ are shown with
  black, gray, and white lines, respectively. Expansion velocity is
  $\approx 50,000~\kms$ at distance $5\times 10^{11}$~cm from the
  ejecta center. Only the innermost part of computational grid is
  shown.}
\label{f:dfd1_ocho}
\end{figure*}

The density and compositional structure of the homologously expanding
model ejecta are shown in Fig.~\ref{f:dfd1_ocho}.
%
%
%
%
%
%
Here we show only the innermost part of the grid with expansion
velocities reaching $\approx 50,000~\kms$ at $r=5\times
10^{11}$~cm. The remaining part of the volume contains a low-density
shocked ambient medium and the supernova shock. The bulk of the ejecta
material displays essentially the same structure as in early
post-detonation models discussed in the previous section. Most visible
differences can be found in the outermost regions that here were
already swept by the shock. The composite character of the bulk of the
ejecta is preserved with a featureless core rich in iron group
elements and the outer strongly mixed layers rich in silicon group
elements. The compositional dichotomy of the outer layers also
reflects the contribution of two processes to the explosion. The inner
well-defined silicon-rich ring also contains an admixture of calcium
(shown with white isocontours in Fig.~\ref{f:dfd1_ocho}). This is due
to the detonation wave nucleosynthesis calculation done with {\tt
aprox13t} nuclear network. The outer silicon-rich and rather
fragmented shell is devoid of calcium as this species was not
considered in the approximate 3-stage flame burner. Improving upon the
approximate nucleosynthesis is one of the urgent future tasks,
especially given the importance of the outer ejecta layers in
formation of supernova spectrum \citep{kasen+05}. It is also
interesting to note that the each deflagration region seems
responsible for forming its own outer silicon-rich ring. This is
evident in model Y100YM25 (Fig.~\ref{f:dfd1_ocho}(c)).

Table~\ref{t:dfd_ho}
%
%
%
\begin{table*}
\caption{Homologous DFD models\tablenotemark{a}}\label{t:dfd_ho}

\begin{tabular}{llllllll}
Model          &  Y12          &  Y25          &  Y50          &  Y100         &  Y75YM25      &  Y100YM25     &  Y75YM50     \\
\tableline
$E_t$          &  1.357        &  1.496        &  1.515        &  1.516        &  1.464        &  1.384        &  1.075       \\
$E_i$          &  1.59\ee{-4}  &  8.38\ee{-5}  &  7.15\ee{-5}  &  7.09\ee{-5}  &  5.34\ee{-4}  &  2.87\ee{-5}  &  1.97\ee{-3} \\
$-E_p$         &  2.52\ee{-3}  &  2.39\ee{-3}  &  2.38\ee{-3}  &  2.38\ee{-3}  &  2.31\ee{-3}  &  2.30\ee{-3}  &  2.56\ee{-3} \\
\nuc{4}{He}    &  8.03\ee{-3}  &  1.13\ee{-2}  &  1.15\ee{-2}  &  1.10\ee{-2}  &  1.03\ee{-2}  &  8.36\ee{-3}  &  2.25\ee{-3} \\
\nuc{12}{C}    &  8.73\ee{-3}  &  5.49\ee{-3}  &  3.30\ee{-3}  &  4.56\ee{-3}  &  1.29\ee{-2}  &  2.05\ee{-2}  &  2.52\ee{-2} \\
\nuc{16}{O}    &  0.107        &  4.65\ee{-2}  &  4.48\ee{-2}  &  3.91\ee{-2}  &  7.54\ee{-2}  &  9.82\ee{-2}  &  0.237       \\
\nuc{20}{Ne}   &  4.41\ee{-4}  &  3.79\ee{-4}  &  3.28\ee{-4}  &  4.78\ee{-4}  &  1.04\ee{-3}  &  9.53\ee{-4}  &  9.73\ee{-4} \\
\nuc{24}{Mg}   &  8.70\ee{-2}  &  3.40\ee{-2}  &  3.42\ee{-2}  &  2.81\ee{-2}  &  4.51\ee{-2}  &  6.74\ee{-2}  &  0.194       \\
\nuc{28}{Si}   &  0.127        &  7.28\ee{-2}  &  6.07\ee{-2}  &  5.74\ee{-2}  &  8.00\ee{-2}  &  0.137        &  0.202       \\
\nuc{32}{S}    &  7.03\ee{-2}  &  3.65\ee{-2}  &  3.06\ee{-2}  &  3.18\ee{-2}  &  4.21\ee{-2}  &  8.75\ee{-2}  &  0.124       \\
\nuc{36}{Ar}   &  1.64\ee{-2}  &  8.26\ee{-3}  &  6.91\ee{-3}  &  7.36\ee{-3}  &  3.97\ee{-3}  &  2.07\ee{-2}  &  2.95\ee{-2} \\
\nuc{40}{Ca}   &  1.82\ee{-2}  &  8.95\ee{-3}  &  7.53\ee{-3}  &  8.09\ee{-3}  &  1.02\ee{-2}  &  2.20\ee{-2}  &  3.24\ee{-2} \\
\nuc{44}{Ti}   &  1.41\ee{-5}  &  9.35\ee{-6}  &  3.02\ee{-5}  &  1.35\ee{-5}  &  2.71\ee{-5}  &  2.71\ee{-5}  &  2.58\ee{-5} \\
\nuc{48}{Cr}   &  2.96\ee{-4}  &  1.49\ee{-4}  &  1.42\ee{-4}  &  1.42\ee{-4}  &  1.78\ee{-4}  &  3.43\ee{-4}  &  4.83\ee{-4} \\
\nuc{52}{Fe}   &  6.50\ee{-3}  &  3.43\ee{-3}  &  3.01\ee{-3}  &  2.91\ee{-3}  &  3.49\ee{-3}  &  6.85\ee{-3}  &  1.03\ee{-2} \\
\nuc{56}{Ni}   &  0.926        &  1.147        &  1.173        &  1.186        &  1.075        &  0.895        &  0.510       \\

\end{tabular}
\tablenotetext{a}{Total energy, $E_t$, and potential energy, $E_p$, in
units $10^{51}$~erg; isotopic yields in solar masses.}

\end{table*}
%
%
%
presents approximate nucleosynthetic yields and final explosion
energies for the complete set of homologous DFD models. The homologous
character of the models is confirmed by the consistently small
fraction of potential and internal energies as compared to the total
energy. In most models, explosion energies are in the range $1.3-1.5$
foe. These produce between $0.9$ to almost $1.2$ solar masses of
\nuc{56}{Ni} and between $0.1$ and $0.24$ solar masses of intermediate
elements. Although model nickel masses may appear relatively
high at first, such high nickel masses might be typical for
significant fraction of objects \citep{stritzinger+06}.
Furthermore, our estimates of nickel mass are likely the upper limits
given {\tt aprox13t} nucleosynthesis does not account for production
of other iron-group elements, e.g.\ stable isotopes like \nuc{54}{Fe}.
Model Y75YM50 is the least energetic ($E_t\approx 1.08\times
10^{51}$~erg), produces the least amount of nickel ($\approx
0.51~\msun$) and more than a half solar mass of intermediate mass
elements. As we discussed earlier, we believe that this model should
not be considered as a DFD, as the shock to detonation transition was
likely being promoted by numerics.
\section{Discussion}\label{s:discussion}
We studied the fate of a massive carbon/oxygen white dwarf following
an off-center mild ignition. We found that such initial configurations
do not produce direct explosions. Only a small amount of stellar fuel
is initially consumed and the released energy is used to expand the
progenitor. This is in agreement with several previous independent
studies in which the deflagration was either intrinsically weak
\citep{arnett+94_2dddt} or was initiated slightly off-center
\citep{niemeyer+96,livne+05}.

We found that the following evolution of the perturbed stellar
material leads to the formation of isolated wave-dominated regions
inside unburned material.  We considered these regions capable of
launching a detonation wave through a shock-to-detonation
transition. We observed the resulting detonations eventually consuming
the bulk of the unburned progenitor. These detonating failed
deflagrations are energetic events with explosion energies $\approx
1.3-1.5$ foe.

The model DFD ejecta appear composite, reflecting the presence of two
different physical processes contributing to the explosion. The
central parts of the ejecta are composed of a mildly deformed but
completely featureless central core rich in iron peak
elements. Stronger deformations may require different physics, e.g.\
rotation \citep{hoeflich05}.  The core is surrounded by a slightly
inhomogeneous inner ring rich in silicon group elements, a product of
the detonation burning at low densities. Finally, the outermost layer
is highly turbulent containing a mix of deflagration products and
unburned material.  Preliminary nucleosynthesis results indicate that
DFD models typically produce over $0.9~\msun$ of iron group elements
and $0.3\msun$ intermediate elements. The burn is almost complete
leaving essentially no carbon.

Our conclusions are based on calculations using a revised numerical
scheme that contains substantial improvements.  We found that the
energetics of deflagration stage originally considered in
\citep[PCL]{calder+04} tended to overestimate buoyancy effects by a
factor $\approx 3$. We used a set of self-heating nuclear network
calculations and implemented a density-dependent energy release
scheme. Additional modifications to the approximate nucleosynthesis
were included to improve the dynamics of the early phases still
further. With the revised energetics, we calibrated the numerical
flame speed to match the results of detailed calculations of
\cite{timmes+92}.

The revised deflagration code was subsequently verified against an
independent set of results of centrally ignited deflagrations obtained
with the Garching supernova code. Good qualitative agreement was found
between the two codes. The database of computer models is offered
on-line to facilitate future verification (code-to-code comparison)
studies.

Furthermore, a numerical procedure to stabilize model progenitors has
been developed. These stabilized progenitors not only provided initial
conditions for supernova simulations but also allowed us to examine
the fidelity of hydrodynamic advection in axisymmetric situations. In
particular we found that, on the one hand, no numerically stable
progenitors can be obtained if resolution is too low and, on the other
hand, small perturbations are strongly amplified near the symmetry
axis in highly resolved models.  This analysis allowed us to identify
the optimal resolution for our supernova calculations.

We analyzed the observed pattern of shock-to-detonation transitions
(SDTs) in some detail. We identified the presence of sufficiently
dense fuel, strongly kinetic flow, wave formation, and a persistent
confinement of the region with additional pressure increase due to
nuclear burning in the shocked gas as necessary conditions for a
SDT. We found that the phenomenon of SDT is not exclusively associated
with the presence of a symmetry axis. We also found that SDTs can
occur in regions completely free of possible geometrical boosting
effects, i.e.\ near the equatorial plane.

However, transitions to detonations were not a robust prediction of
such models. There are some possible reasons for that. For example, by
assuming axisymmetry we eliminate an angular direction in which
additional perturbations may develop. Such perturbations will enrich
the flow field creating more seed points for SDT and, at the same
time, increase the amplitude of fluctuations.  That is, the assumed
symmetry is likely limiting the possible wave interactions and
presumably denying extreme events such as a SDT. In addition, for
stability reasons, our calculations had to be performed using
suboptimal numerical resolution. This caused a strong numerical
damping and limited sampling of the perturbed regions harming chances
for observing SDTs still further.  On the other hand, we did not
include physics that may, effectively, make the system behavior appear
more viscous (e.g.\ magnetic fields) inhibiting formation of small
scale structures.

Our findings also hinge on the assumption that the SDT process is
relatively insensitive to details of evolution on scales unresolved in
our simulations. This remains to be demonstrated, ideally in the
course of dedicated highly-resolved model calculations of
compressionally heated fuel-rich degenerate mixtures. It will be a
daunting task. If any parallel can be drawn, experience accumulated by
modelers of inertially confined fusion systems might be of great help
in such studies \citep{atzeni+04,drake06}. Even if such models are
successfully constructed, many doubts will remain regarding the
outcome of such calculations given how limited our knowledge about
real systems is. For example, as we mentioned before (PCL), although
we consider a pure carbon/oxygen progenitor it is almost certain that
in nature progenitor's surface layers contain admixture of helium
\citep{nomoto82,yoon+05b}. Compositional changes will affect
energetics of nuclear burn adding entirely new dimension to the
problem.

Being mindful of numerous simplifying assumptions and model
inaccuracies, the essential findings of this work are, therefore,
rather modest and can be summarized as evidence of strong, localized,
and prolonged shock heating in regions rich in fuel. We note that
these are necessary conditions for shock-to-detonation transition to
occur. We believe this observation is independent of particular
details of our model, especially numerics, making SDT one of prime
suspects for triggering detonations in SN~Ia.

However, even if no seed point forms a detonation through SDT, this
second, after the initial deflagration, failure to unbind the star in
no way automatically implies supernova will not occur. Perhaps just
the opposite. The extensive large-scale mixing of deflagration
products with unburned outer stellar layers combined with abundantly
present strong acoustic perturbations appear the conditions are ripe
for the Zeldovich gradient mechanism \citep{khokhlov+97}. We may
expect that for rotating progenitors \citep{yoon+05a} perturbations of
surface layers will be even stronger due to presence of additional
shear component. If all these opportunities are missed, the white
dwarf might still be given another chance to produce the supernova. A
failed attempt to explode would then be a beginning of a cycle that
repeats, perhaps several times, as the expanded white dwarf
eventually cools down, shrinks, and prepares for another
ignition. That is, the explosion process might be a lengthy one, a
kind of laborious slow death.

As the observations improve, we are also beginning to collect evidence
that the degree of diversity of SN~Ia might be greater than original
anticipated (or desired!). Recent observations of the peculiar
supernova SN~2002cx by \cite{jha+06_2002cx} argue in favor of low
energy explosion and large degree of mixing in this object, two
characteristics that essentially preclude any involvement of a
detonation in the explosion process. Other objects listed by
\cite{jha+06_2002cx} may belong to SN~2002cx class. These rare
peculiar supernovae might be genuine examples of pure
deflagrations. Or perhaps these are objects that underwent several
failed deflagrations leaving only small amount of material to fuel the
detonation. If so, normally bright supernovae might be DFDs that
succeeded early, with the occurrence of a detonation (or lack thereof)
being the primary element determining the observational properties of
a given event.

On the other hand, SNe~Ia display some characteristics that we find
difficult to explain in the framework of pure deflagrations. One
example are iron-rich clumps found in Type~Ia SNRs
\citep{warren+04,warren+05}. In the model proposed here, the inner
ring of intermediate elements seems to be a natural site for producing
nickel-rich blobs that may float away from the central core
\citep{blondin+01,wang+01}. Those radioactive blobs may also be
explained in pure deflagration models that naturally produce clumpy
ejecta. However, dominant and isolated regions rich in iron-group
elements like the one observed in Tycho SNR \citep{vancura+95,warren+05} can
hardly be produced in a pure deflagration in which several such
regions are expected to be simultaneously present. In DFDs, such an
isolated clump located near the outer edge of the supernova remnant
might be a material burned deep in the progenitor core and transported
to the stellar surface by one of the deflagrating bubbles.

We cannot address the above question without detailed nucleosynthesis
calculations. This is one of the possible future directions for
research. In addition, the relaxation of the assumption of axial
symmetry, although costly, will be necessary. But even with only the
current approximate nucleosynthesis and simplified geometry of the
problem, we are in a position to conduct preliminary validation
studies against observations for a subset of DFD models. This will be
the subject of the next communication in this series.
\acknowledgements
Todd Dupont and Dan Kasen provided me with both support and
encouragement for continuing this work.  I would like to thank Timur
Linde for contributing the flame surface integrator, Frank Timmes for
providing the initial white dwarf model, Bronson Messer for helping in
verifying and developing the approximate deflagration network, Cal
Jordan for extending the original nuclear burning network, and the
anonymous referee for comments that led to the improvements of the
initial version of the paper. I enjoyed stimulating and helpful
discussions with Carles Badenes, Peter H\"oflich, Dan Kasen, Alexei
Khokhlov, Jens Niemeyer, and Joe Shepherd. This work is supported in
part by the U.S.\ Department of Energy under Grant No.\ B523820 to the
Center for Astrophysical Thermonuclear Flashes at the University of
Chicago. It benefited from the INCITE award and later computing
allocations provided by the National Energy Research Scientific
Computing Center, which is supported by the Office of Science of the
U.S.\ Department of Energy under Contract No.\
DE-AC03-76SF00098. Additional computations were performed on the
Teraport cluster, part of the Teraport project at the University of
Chicago funded through National Science Foundation Grant No.\ 0321253.

\end{document}